\documentclass[pra,aps,letterpaper,twocolumn]{revtex4}

\usepackage{hyperref}
\usepackage{graphicx}
\usepackage{epsfig}
\usepackage{float}

\begin{document}
\title{Computation by measurements: a unifying picture}

\author{Panos Aliferis}
\email[email: ]{panos@caltech.edu}

\affiliation{Institute for Quantum Information, Caltech, MC 107-81, Pasadena, CA 91125, USA}

\author{Debbie W. Leung}
\email[email: ]{wcleung@caltech.edu}

\affiliation{Institute for Quantum Information, Caltech, MC 107-81, Pasadena, CA 91125, USA}

\date{\today}

\begin{abstract}
The ability to perform a universal set of quantum operations based solely on static resources and measurements presents us with a strikingly novel viewpoint for thinking about quantum computation and its powers. We consider the two major models for doing quantum computation by measurements that have hitherto appeared in the literature and show that they are conceptually closely related by demonstrating a systematic local mapping between them. This way we effectively unify the two models, showing that they make use of interchangeable primitives. With the tools developed for this mapping, we then construct more resource-effective methods for performing computation within both models and propose schemes for the construction of arbitrary graph states employing two-qubit measurements alone.

\end{abstract}

\maketitle

\raggedbottom

\section{Introduction}
Faced with the question of describing a quantum computation in terms of elementary operations, one is almost invariably tempted to answer by drawing lines signifying qubits and little boxes signifying unitary operators performed on them.  Thus quantum computation is usually viewed as some more or less complicated manipulation of the initial quantum state, the sum total and ultimate goal of which is to apply a certain unitary operator on it.  Measurement naturally appears at the very end of the computation, and is generally considered harmful for the coherence if it is included in the main body of the computation, due to its inherent irreversibility. In this respect, the computational power bestowed on us by quantum computers appears to depend vastly on our ability to perform unitary operations on our qubits, postponing any measurements until the very end when the result of the computation is ready to be read off. In fact, the standard model of quantum computation \cite{divincenzo95} consists of preparing a standard initial state $|0\rangle^{\otimes n}$, applying an arbitrary unitary transformation and performing measurements in the very end. 

The first indication that measurements can be an integral part of the main body of our quantum computation was given by the fault-tolerant constructions for the $\pi$/8 and the Toffoli gates \cite{boykin99,shor96,zhou00}.  Both of these make use of measurements and special ancillary states for the fault-tolerant implementation of the gates, with the ancillary states in turn prepared by measurements. But trully, the ability to perform universal quantum computation based on measurements alone was not fully realized until recently. Two explicit models for doing computation by measurements will be considered: the first one based on one-qubit measurements on a cluster state (1WQC) \cite{raussen01}, and the second one based on two-qubit measurements alone (TQC) \cite{gottesman99,nielsen01,leung01}. Proofs for the universality of both of these models were obtained by reduction to the standard model: preparation of a standard initial state, the ability to perform any unitary operation with arbitrary accuracy and the ability to perform measurement, which is a natural constituent of them both. The ability to realize any unitary transformation was in turn reduced to proving that all gates from a universal set of gates (typically taken to be the one consisting of all one-qubit gates and the {\sc cnot} \cite{barenco95}) can be realized within each model.  

\vspace{0.0cm}
\subsection{One-way quantum computer (1WQC)}
In the 1WQC model, quantum computation is performed on qubits arranged in a regular lattice and prepared initially in a specific entangled state, known as the \emph{cluster state}. Any desired computation is then encoded as a sequence of projective one-qubit measurements of these lattice qubits along certain bases. Although the intermediate measurement results are random, by monitoring the measurement outcomes one is able to exploit the quantum correlations and readapt the future measurement bases in order to effectively steer forward the desired computation. Thus an arbitrary trajectory in the Hilbert space of the input state can be achieved, as quantum information is made to travel within the lattice from the measured qubits to their neighboring ones and thereover until the completion of the measurement sequence.

Conceptually, the easiest way to describe the cluster state is by giving its stabilizer generators. For each qubit, viewed as a lattice site in a lattice $\cal{L}$, the stabilizer consists of generators of the form
\[
    \label{cluster stabilizer}
    K^{(i)}=X^{(i)}\bigotimes\limits_{j \in \text{nbhd}(i)}Z^{(j)} ,
\hspace{0.2cm} \forall i \in \cal{L} ,
\]
where nbhd$(i)$ is the set of qubits in the neighborhood of the qubit $i \in \cal{L}$ and $\{I,  X\equiv \sigma_x,  Y\equiv \sigma_y,  Z\equiv \sigma_z\}$ is a notation that will be used in this paper alternatively with the notation $\{\sigma_0, \sigma_1, \sigma_2, \sigma_3\}$ for the Pauli matrices.  The cluster state is a particular example of a family of states known as {\em graph states}, which have stabilizer generators of the same form as the cluster state, with the exception that the vertices are not necessarily viewed as points on a lattice and the neighboring relations are generally given by an adjacency matrix.

Operationally, any graph state can be prepared by various methods. A simple realization is obtained by examining the stabilizer: all lattice qubits need to be initialized in the state $|+\rangle$, creating $\bigotimes_{i \in \cal{L}}|+\rangle$, and then a controlled-phase gate needs to be applied between all pairs of neighboring qubits. The controlled-phase, $\Lambda(Z)$, in the computational basis takes the simple diagonal form
\[
\Lambda(Z) = \left( \begin{array}{cccc}
              1 & 0 & 0 & 0 \\ 0 & 1 & 0 & 0 \\ 0 & 0 & 1 & 0 \\ 0 & 0 & 0 & -1
           \end{array}
             \right).
\]
These controlled-phase gates have the convenient properties that each acts symmetrically between the two qubits on which it is applied and also they all commute with one another.  Physically, the preparation of the graph state can be accomplished by applying the nearest-neighbor homogeneous Ising $\sigma_z\otimes \sigma_z$ interaction on the state $\bigotimes_{i \in \cal{L}}|+\rangle$ for the appropriate period of time and then correcting the resulting state with local unitary operations.

Returning to the issue of how an arbitrary unitary operation can be realized on an already prepared cluster state, we recall that the requirement has been relaxed to just being able to realize all operators from a universal gate set. To this direction, one way to prove the universality of the 1WQC model is to consider disjoint regions of the cluster and use each such region for simulating a certain quantum gate from our selected universal set. Identifying the input qubits of one such region with the output qubits of the previous, an arbitrarily large succession of quantum operations can then be realized.

 
Disjoining lattice regions can be accomplished by selectively disentangling qubits from the cluster state, thus effectively deleting them from the initial lattice. A quick inspection of the stabilizer in fact shows that measuring a qubit in the computational basis is sufficient for this deletion. Indeed, a measurement of the $i^{\rm th}$ qubit in the computational basis corresponds to a measurement of the operator $Z^{(i)}$, which commutes with all stabilizer generators except for that one having the measured qubit as the correlation center, namely $X^{(i)}\bigotimes_{j \in {\rm nbhd}(i)}Z^{(j)}$. Thus measuring $Z^{(i)}$ removes the generator $X^{(i)}\bigotimes_{j \in {\rm nbhd}(i)}Z^{(j)}$ from the stabilizer, while leaving all other stabilizer generators unaltered up to the removal of the measured qubit.

\begin{figure}[htb]
    \begin{center}
      \begin{tabular}{cc}
                         \footnotesize{(a) $U_x(\phi)=e^{-i{\phi \over 2} \sigma_x}$} \hspace{0.4cm} & \footnotesize{(b) $U_z(\theta)=e^{-i{\theta \over 2} \sigma_z}$} \vspace{0.3cm}
                         \\
                         \epsfig{file=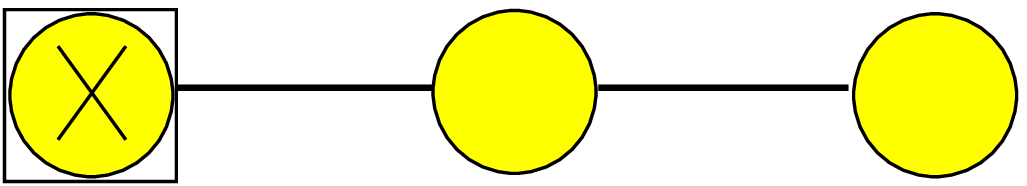,width=3.7cm} 
                                                            \put(-2.05,0.25){$\pm \phi$}
                                                            \put(-3.45,0.8){1}  \put(-1.9,0.8){2}            \put(-0.45,0.8){3}
                                                            \hspace{0.4cm} &
                         \epsfig{file=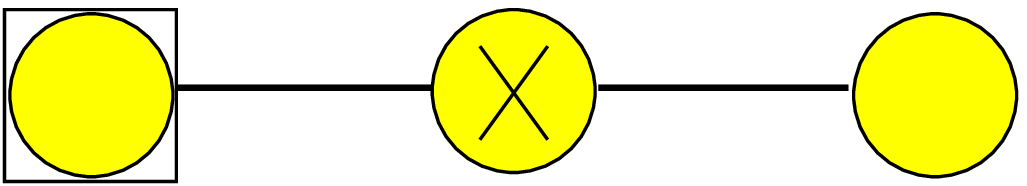,width=3.7cm} \put(-3.6,0.25){$-\theta$}
                                                            \put(-3.5,0.8){1}          \put(-2.0,0.8){2}     \put(-0.45,0.8){3}
                         \\ & \vspace{0.3cm} \\
                         \hspace{-2.5cm} \footnotesize{(c) {\sc cnot}} & \hspace{-3cm}
                                            \parbox{3.7cm}{
                                                           \epsfig{file=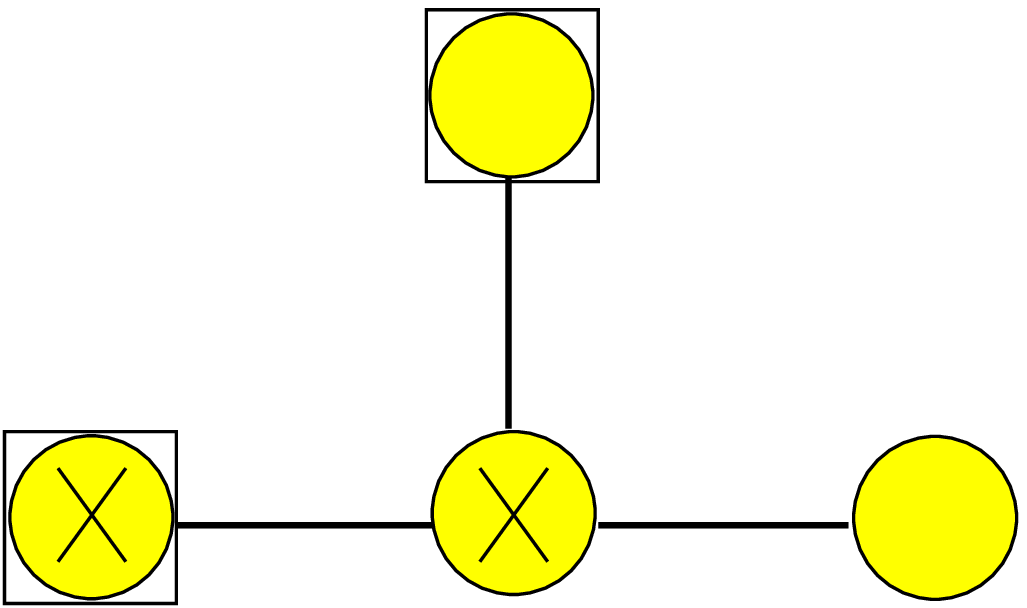,width=3.7cm}
                                                          }
                                                            \put(-3.48,-1.3){1} \put(-1.95,-1.3){2}   \put(-0.43,-1.3){3}   \put(-1.4,0.8){4}
                                                            \put(-4.8,-0.75){\parbox{1cm}{\footnotesize{target in}}} \put(0,-0.75){\parbox{1cm}{\footnotesize{target out}}}
                                                            \put(-3.2,0.8){\footnotesize{control}}
      \end{tabular} \vspace{0.3cm}
\caption{\label{cluster gates}
         \footnotesize{
                       Measurement patterns that realize (a) a rotation of $\phi$ about the $x$-axis, (b) a rotation of $\theta$ about the $z$-axis, (c) the {\sc cnot}$_{4\rightarrow1}$ \cite{raussen01}. A general rotation with Euler decomposition $U_z(\psi)U_x(\theta)U_z(\phi)$ can be built by composing $x$- and $z$-rotations using a total of five qubits \cite{raussen03}. Boxed circles indicate the input qubits. The measurement bases are shown as X, for measurements of the observable $\sigma_x$ corresponding to a projection along the $x$-axis, or as the appropriate angle, $\omega$, with respect to the $x$-axis in the equator of the Bloch sphere for measurements of the observable ${\rm cos}(\omega)\sigma_x+{\rm sin}(\omega)\sigma_y$. The choice between positive or negative angles is made based on the outcomes of previous measurements.
                       }
         }
    \end{center}
\end{figure}

To complete the proof that performing one-qubit measurements on the cluster state is universal for quantum computation, in Fig.$\,$\ref{cluster gates} we explicitly show how certain elementary operations can be realized by one-qubit measurements on the appropriate qubit configurations \cite{raussen03}. As already stated, universality then follows from our ability to simulate any operator from the universal set of one-qubit rotations and {\sc cnot}.

\vspace{0.2cm}

At this point it should be noted that computation is done up to local Pauli corrections, meaning that the quantum state of the qubits at the output will be of the general form $\bigotimes_{i=1}^N \sigma_j^{(i)}U|\Psi\rangle$, where $U$ is the unitary to be applied to the input state $|\Psi\rangle$ and $\sigma_j^{(i)}$, $j=0,\dots ,3$ is one of the Pauli operators applied to the $i^{th}$ of the N output qubits. Additionally, it is important to emphasize that the operation of the 1WQC always starts with a cluster state, and no other input states or ancillae are being used.

\vspace{0.0cm}
\subsection{Teleportation-based quantum computer (TQC)}
The core idea of this scheme lies in the realization that we can modify the basic teleportation protocol \cite{bennett93} in order to also affect a unitary transformation while teleporting a quantum state from one qubit to another. In the context of the TQC, teleportation is therefore viewed as a way of affecting unitary transformations, with its function as simulating a quantum communication channel becoming irrelevant in the absence of distant communicating parties.  

\begin{figure}[htb]
    \begin{center}
      \begin{tabular}{cc}
                         \footnotesize{(a)} \hspace{0.1cm} & \parbox{4cm}{
                                                           \epsfig{file=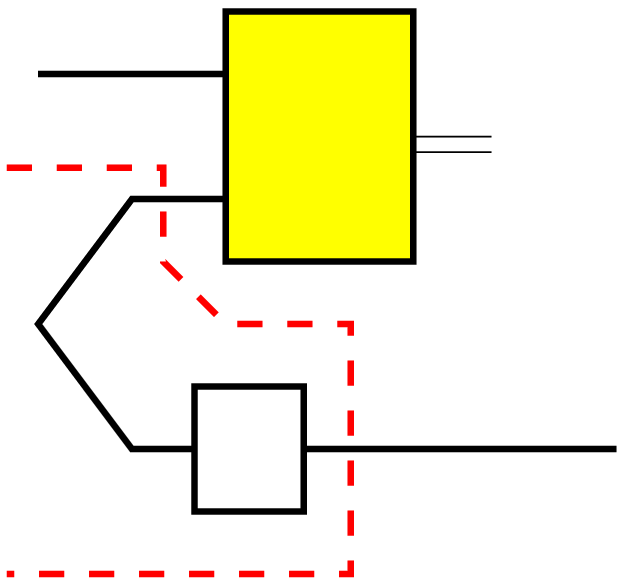,width=3.35cm}
                                                          } \put(-4.05,1.2){$|\Psi\rangle$} \put(-2.25,0.65){\huge{\bf B}}
                                                            \put(-0.95,0.75){$j$}
                                                            \put(-2.45,-0.85){$U$} \put(-0.15,-0.85){$U\sigma_j|\Psi\rangle$}
                         \\ & \\
                         \footnotesize{(b)} \hspace{0.1cm} & \parbox{4cm}{
                                                           \epsfig{file=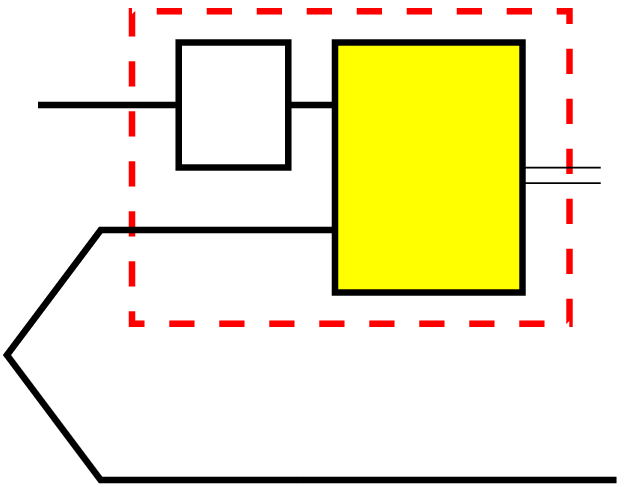,width=3.2cm}
                                                                         } \put(-4,0.75){$|\Psi\rangle$} \put(-2.5,0.7){$U$}
                                                                           \put(-1.7,0.25){\huge{\bf B}}
                                                                           \put(-0.35,0.35){$j$}
                                                                           \put(-0.2,-1.2){$\sigma_jU|\Psi\rangle$}
      \end{tabular} \vspace{0.2cm}
\caption{\label{teleportation}
         \footnotesize{
                       To apply the single-qubit unitary $U$, we either (a) prepare the ancilla $(I\otimes U)|\Phi_0\rangle$ and measure in the Bell basis $\{|\Phi_j\rangle\}_j$ or (b) use a Bell state $|\Phi_0\rangle$ and perform the generalized Bell measurement in the basis $\{(U^\dagger \otimes I) |\Phi_j\rangle\}_j$. Lines joined on one end indicate a Bell pair $|\Phi_0\rangle$ throughout this paper.
                       }
         }
   \end{center}
\end{figure}

In Fig.$\,$\ref{teleportation} we describe two alternative ways for applying an arbitrary one-qubit unitary $U$ to the input quantum state $|\Psi\rangle$ using only two-qubit measurements: either apply the unitary to the state after it has been teleported, which can equivalently be viewed as absorbing $U$ directly into the special ancilla $(I\otimes U)|\Phi_0\rangle$ before performing the Bell measurement \cite{gottesman99,nielsen01}, or apply the unitary to the input state before teleporting and combine it with the measurement to form a generalized Bell measurement in the basis $\{(U^\dagger \otimes I)|\Phi_j\rangle\}_j$ \cite{leung01,leung03}. A Bell measurement is conventionally defined as the complete two-qubit measurement along the basis $\{|\Phi_j\rangle\}_j$, with $j=0,\dots ,3$, where
\[
|\Phi_{0,3}\rangle={|00\rangle \pm |11\rangle \over \sqrt{2}} ,\hspace{0.2cm} |\Phi_{1,2}\rangle={|01\rangle \pm |10\rangle \over \sqrt{2}}.
\]

\begin{figure}[tb]
     \begin{center}
         \hspace{-1.5cm} \epsfig{file=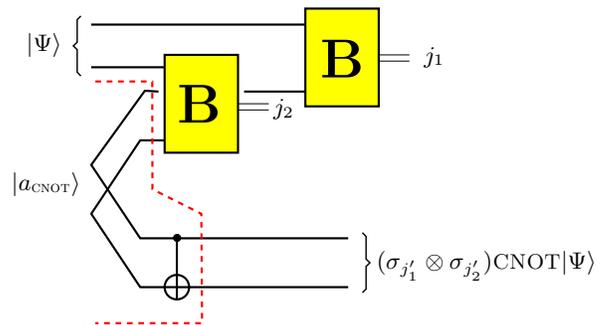, width=4.5cm} \put(-0.45,0.75){$(\sigma_{j_1'}\otimes \sigma_{j_2'})\text{{\footnotesize \sc CNOT}}|\Psi\rangle$}
                                                \put(-5.1,3.65){$|\Psi\rangle$}
                                                \put(-1.2,3.3){\huge{\bf B}} \put(-3.1,2.7){\huge{\bf B}}
                                                \put(0.18,3.5){$j_1$} \put(-1.82,2.8){$j_2$}
                                                \put( -5.3,1.8){$|a_{\text{{\sc cnot}}}\rangle$}
\caption{\label{CNOT}
         \footnotesize{
                       The ancilla $|a_{\text{{\sc cnot}}}\rangle$ is prepared separately using two-qubit measurements alone \cite{leung01}. Subsequently two Bell measurements are performed, one for each input qubit. {\sc cnot} being in the Clifford group allows us to commute the Pauli corrections through it and write $\text{{\sc cnot}}(\sigma_{j_1}\otimes \sigma_{j_2})=(\sigma_{j_1'}\otimes \sigma_{j_2'})\text{{\sc cnot}}$.
                       }
         }
      \end{center}
\end{figure}

In order to form a universal set of operators we need to augment our set containing all one-qubit rotations with one more two-qubit operator, the {\sc cnot}. The {\sc cnot} (as well as any other two-qubit unitary) can also be simulated in the TQC model by doubling the number of ancillary qubits and Bell measurements. The construction, employing the ancilla $|a_{\text{{\sc cnot}}}\rangle$, is shown in Fig.$\,$\ref{CNOT}.

Although both circuits in Fig.$\,$\ref{teleportation} were presented as extensions to the basic teleportation scheme, an important trade-off between them should be stressed \cite{leung03}. When the unitary to be realized is not in the Clifford group (defined as the normalizer of the Pauli group \cite{book}), the circuit of Fig.$\,$\ref{teleportation}a implements $U$ non-deterministically: commuting the Pauli correction $\sigma_j$ through $U$ can, for certain values of $j$, result in realizing a totally different unitary.  Extra teleportation steps would be needed (each to undo the faulty gate and reattempt the intended unitary gate $U$) until the outcome of the Bell measurement is zero, indicating the successful application of the gate. Hence, the complication of a nondeterministic number of teleportation steps and the additional complexity of the ancillae are traded for the simplicity of the same Bell measurement for any unitary. On the other hand, the circuit of Fig.$\,$\ref{teleportation}b realizes any $U$ deterministically (the Pauli corrections appearing to the left of $U$) with the additional complication of a generalized Bell measurement that must be adapted each time according to which operator is to be implemented. Because of this tradeoff, the circuit of Fig.$\,$\ref{teleportation}b will be used for all single-qubit operations and that of Fig.$\,$\ref{CNOT} for the {\sc cnot}, to ensure that the Pauli corrections occur to the left of the applied gate in both cases.

Thus in the TQC, just as in the 1WQC, we can {\em deterministically} perform the quantum computation up to local Pauli corrections. This is sufficient, since it will just translate into left propagating the Pauli errors in the case when the next applied unitary is in the Clifford group, or appropriately modifying the subsequent measurement bases to compensate for the accumulated Pauli errors up a given point in the case of a non-Clifford single-qubit gate.

In section II we explicitly demonstrate how the two models can be mapped to one another based on the universal set of one-qubit rotations and {\sc cnot}. Using the tools developed for this mapping, in section III we derive a useful circuit that implements the remote-$\Lambda(Z)$ gate and utilize it to propose more resource-effective methods for performing computation in both the TQC and 1WQC models and also to develop schemes for the construction of arbitrary graph states employing a combination of complete and incomplete two-qubit measurements. Graph states have emerged as the natural generalization of the cluster state and are of fundamental importance in the operation of the 1WQC, as they provide the starting point for embarking upon error-correction and fault-tolerance \cite{hein03,nest03}.

\vspace{0.0cm}
\section{Mapping between the 1WQC and the TQC}
To gain some insight into how to establish an equivalence, we will begin by showing how teleportation is realized in the two models. We will then proceed to explicitly show the mapping for arbitrary $x$- and $z$-rotations and for the {\sc cnot}.

\vspace{0.0cm}
\subsection{Teleportation}
\begin{figure}[htb]
              \vspace{0.1cm}
              \epsfig{file=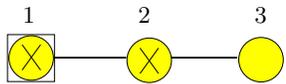,width=3.7cm}
                                                      \put(-3.48,0.8){1}  \put(-1.95,0.8){2} \put(-0.4,0.8){3} 
              \vspace{0.2cm}
\caption{\label{1wqc_tel}
         \footnotesize{The quantum wire in the 1WQC.
                      }
         }
\vspace{0.2cm}  
\end{figure}

In the 1WQC model, a wire for teleporting a quantum state is formed by three qubits connected in the pattern sketched in Fig.$\,$\ref{1wqc_tel}. The first two qubits are measured in the $X$-basis, resulting in the teleportation of the input state from the first to the third qubit.

\vspace{0.2cm}

\begin{figure}[htb]
     \begin{center}
          \begin{tabular}{c}
                            \hspace{-1cm} \footnotesize{(a)} \hspace{0.7cm}
                                            \parbox{4cm}{
                                                         \epsfig{file=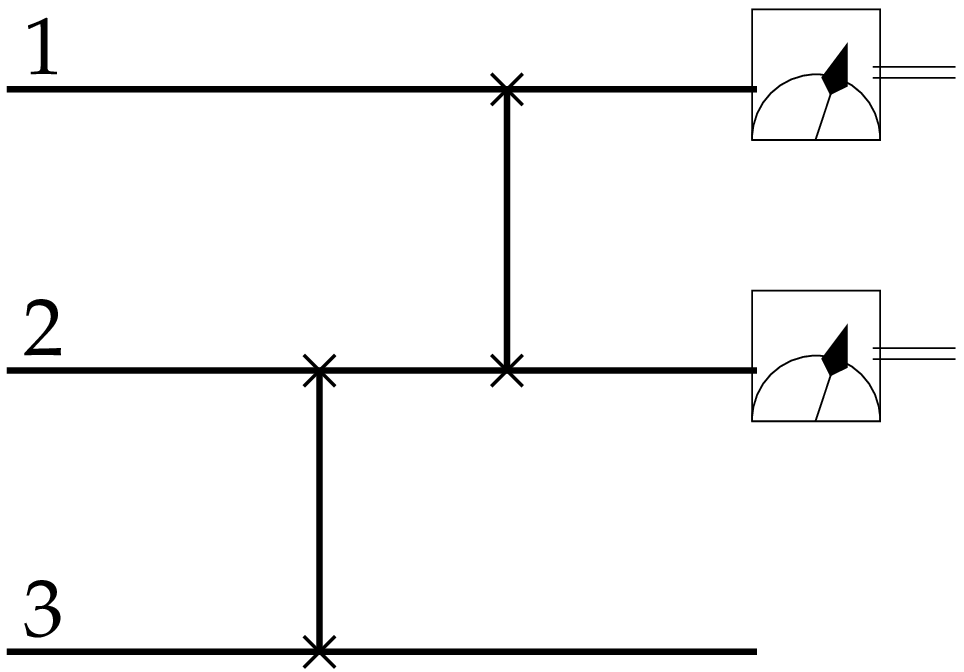,width=4cm}
                                                         } \vspace{0.5cm} \put(-4.6,1.1){$|\Psi\rangle$} \put(-4.6,-0.1){$|+\rangle$} \put(-4.6,-1.3){$|+\rangle$}
                                                                          \put(0,1.1){$j_1$} \put(0,-0.1){$j_2$}
                                                                          \put(-0.35,0.8){{\scriptsize X}} \put(-0.35,-0.4){{\scriptsize X}}
                            \\
                            \hspace{-1.2cm} \footnotesize{(b)} \hspace{0.4cm}
                                            \parbox{4.4cm}{
                                                         \epsfig{file=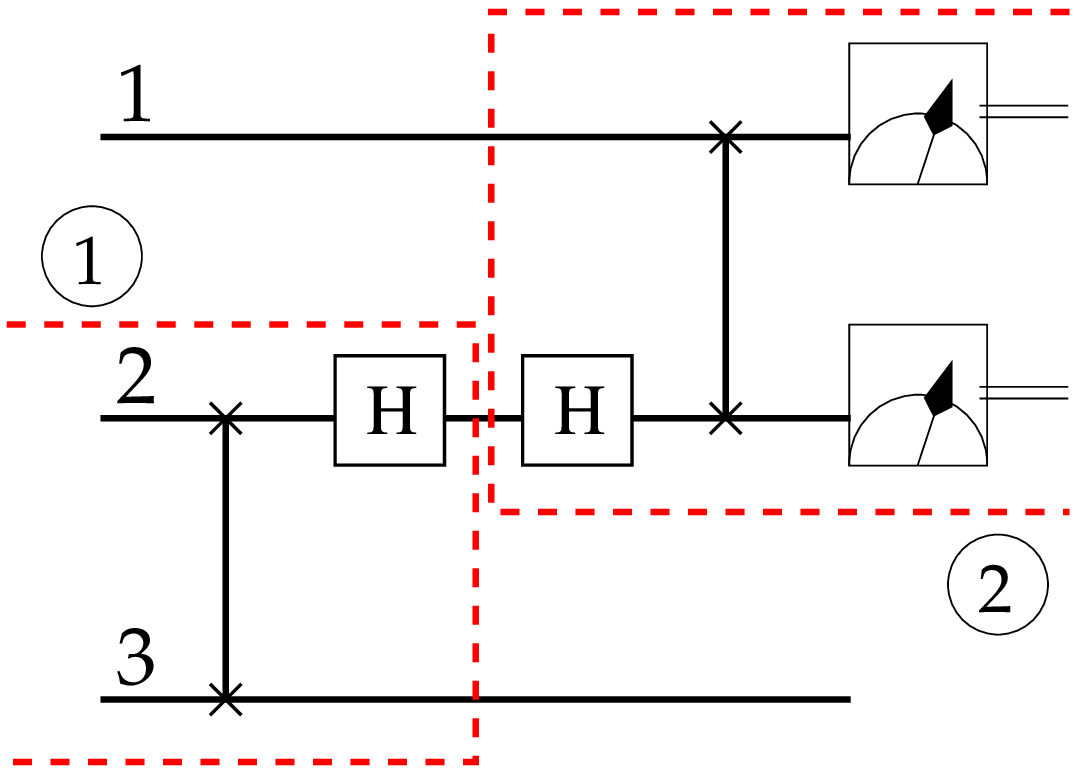,width=4.4cm}
                                                         }\put(-4.55,1.1){$|\Psi\rangle$} \put(-4.55,-0.1){$|+\rangle$} \put(-4.55,-1.2){$|+\rangle$}
                                                          \put(0.1,1.1){$j_1$} \put(0.1,-0.05){$j_2$}
                                                          \put(-0.25,0.8){{\scriptsize X}} \put(-0.25,-0.35){{\scriptsize X}}
         \end{tabular}
\vspace{0.3cm}
\caption{\label{1wqc_tel_network}
         \footnotesize{
                       In this paper, the controlled-phase gate is denoted by a line segment connecting two qubits, with x's drawn on both ends to emphasize that it acts symmetrically between them. Measurement boxes are labelled on their lower right corner by the corresponding measurement basis. To go from (a) to (b) we insert the identity in the form $H^2=I$. Then we identify box 1 as preparing a Bell state $|\Phi_0\rangle$ and box 2 as performing a Bell measurement.
                      }
         }
     \end{center}
\end{figure}

\noindent In all our 1WQC gate diagrams, the effective input qubits (carrying the quantum state from the previous part of the computation) will be distinguished by being drawn boxed. As already explained in the introduction, all other qubits in the lattice are initialized in the $|+\rangle$ state, while the nearest-neighbor interaction that was used to create the cluster state can be viewed as affecting a controlled-phase gate between all pairs of neighboring qubits. An equivalent circuit representation of the quantum wire is therefore that of Fig.$\,$\ref{1wqc_tel_network}a.

The circuit of Fig.$\,$\ref{1wqc_tel_network}b can now be obtained from the one of Fig.$\,$\ref{1wqc_tel_network}a by inserting the identity $H^2=I$. In Fig.$\,$\ref{1wqc_tel_network}b, the stabilizer of the second and third qubits is initially $XI$, $IX$. After conjugation with $(H\otimes I)\Lambda(Z)$ it becomes 
\[
  \begin{array}{ccccc}
        \begin{array}{c}
                     XI \\ IX
        \end{array}               & \stackrel{\Lambda(Z)}{\longrightarrow} & \begin{array}{c}
                                                                                          XZ \\ ZX
                                                                             \end{array} & \stackrel{H\otimes I}{\longrightarrow} & \begin{array}{c}
                                                                                                                                                 ZZ \\ XX
                                                                                                                                    \end{array} 
  \end{array}             
\] \pagebreak

This is the stabilizer of the state $|\Phi_0\rangle$, proving that box 1 prepares a Bell pair. Similarly, in order to interpret box 2, we begin with the stabilizer generators of the output $(-1)^{j_1}X \otimes (-1)^{j_2} X$, where $j_1, j_2 \in \{0,1\}$ are the outcomes of the X-measurements on the first two qubits. Conjugating it backwards through $(I\otimes H)\Lambda(Z)$ we obtain
\[
  \begin{array}{ccccc}
                      \begin{array}{c}
                      (-1)^{j_1} XI \\(-1)^{j_2}IX
                      \end{array} 
                                         & \stackrel{\Lambda(Z)}{\longrightarrow} &
                      \begin{array}{c}
                      (-1)^{j_1}XZ \\(-1)^{j_2}ZX
                      \end{array} 
                                         & \stackrel{I\otimes H}{\longrightarrow} &
                      \begin{array}{c}
                      (-1)^{j_1}XX \\(-1)^{j_2}ZZ
                      \end{array}        
  \end{array}
\]
Box 2 can therefore be viewed as realizing a Bell measurement in the basis $\{|\Phi_j\rangle\}_j$, with $j \in \{0,1,2,3\}$ now given in binary as $j=(j_1,j_1\oplus j_2)$. 

\begin{figure}[htb]
    \begin{center}
      \begin{tabular}{cc}
                         \footnotesize{(a) Bell state preparation} \hspace{0.7cm} & \footnotesize{(b) Bell measurement} \vspace{0.3cm}
                         \\
                         \hspace{0.2cm} \epsfig{file=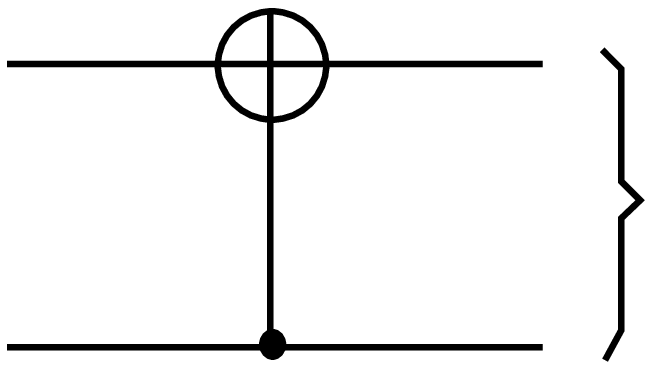,width=2.5cm} \put(-2.95,1.1){$|0\rangle$} \put(-3.0,0.0){$|+\rangle$} \put(0.1,0.6){$|\Phi_0\rangle$}
                                                           \hspace{0.6cm} &
                         \epsfig{file=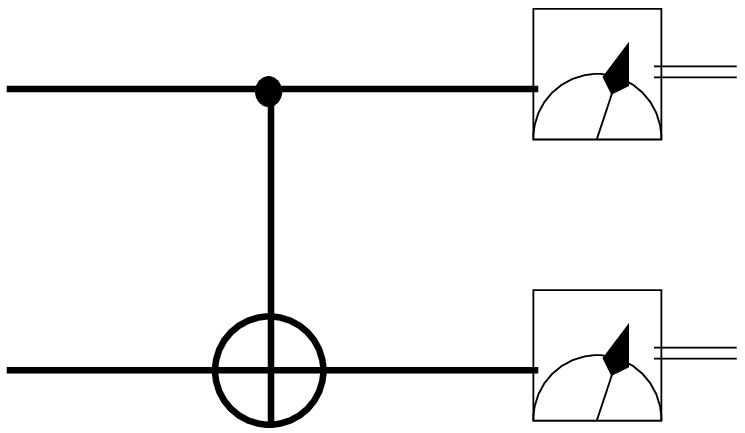,width=2.9cm} \put(0.15,1.2){$j_1$} \put(0.15,0.2){$j_2$}
                                                           \put(-0.25,0.95){{\scriptsize X}}     \put(-0.25,-0.15){{\scriptsize Z}}
                         \\ & \vspace{0.3cm} \\
                         \multicolumn{2}{c}{
                                           \hspace{-0.2cm} \footnotesize{(c)} \hspace{0.6cm}
                                           \parbox{1.25cm}{\epsfig{file=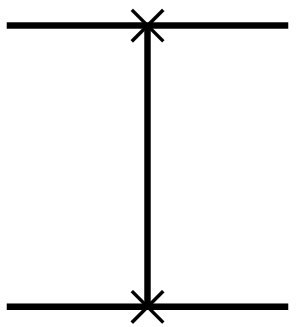,width=1.25cm}} \hspace{0.2cm} \parbox{1cm}{$\Longleftrightarrow$} \hspace{0.2cm}
                                           \parbox{2.6cm}{\epsfig{file=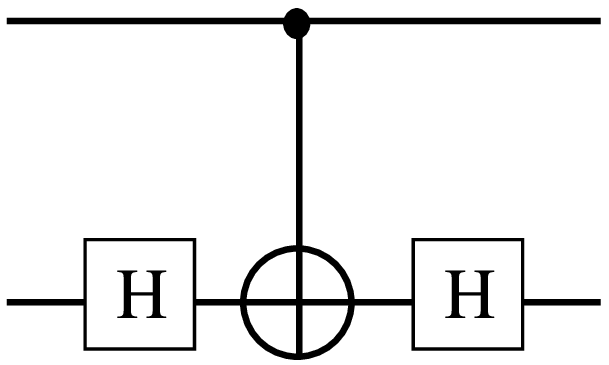,width=2.6cm}}
                                           }
      \end{tabular}
\vspace{0.2cm}
\caption{\label{identities}
         \footnotesize{
                       (a) Simple circuit preparation of the Bell state $|\Phi_0\rangle$, (b) Bell measurement in the basis $\{|\Phi_j\rangle\}_j$, with $j=(j_1,j_1\oplus j_2)$ in binary, and (c) a circuit representation of the identity $\Lambda(Z)=(I\otimes H)\text{CNOT}_{1\rightarrow 2}(I\otimes H)$.
                       }
         }
    \end{center}
\end{figure}

Naturally, the above derivation can also be carried out by working explicitly with states and without any use of the stabilizer formalism. For that purpose, the identities of Fig.$\,$\ref{identities} provide a quick way to visualize the operation of boxes 1 and 2. Indeed, using the identity of Fig.$\,$\ref{identities}c, box 1 is transformed into the Bell state preparation circuit of Fig.$\,$\ref{identities}a and similarly, box 2 is transformed into the Bell measurement of Fig.$\,$\ref{identities}b.

\vspace{0.2cm}

Boxes 1 and 2 then establish the equivalence of the circuit of Fig.$\,$\ref{1wqc_tel_network}b and the teleportation circuit of Fig.$\,$\ref{teleportation} for $U=I$. Overall, the transition from Fig.$\,$\ref{1wqc_tel} to the circuits of Fig.$\,$\ref{1wqc_tel_network} exhibits the mapping from the 1WQC pattern to the well-known teleportation circuit in the TQC. All steps can equally well be traced backwards, proving also the mapping from the TQC to the 1WQC. 

\vspace{0.0cm}
\subsection{Rotation about the $x$-axis}
\begin{figure}[htb]
     \begin{center}
          \begin{tabular}{c}
                            \hspace{-1cm} \footnotesize{(a)} \hspace{0.7cm}
                                            \parbox{4cm}{
                                                         \epsfig{file=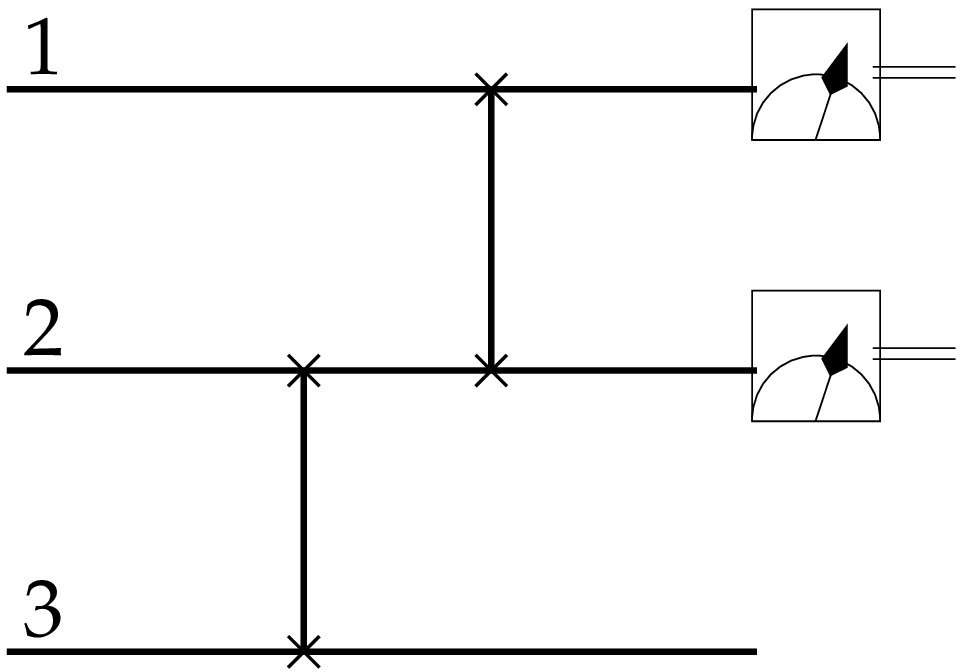,width=4cm}
                                                         } \put(-4.55,1.1){$|\Psi\rangle$} \put(-4.55,-0.1){$|+\rangle$} \put(-4.55,-1.3){$|+\rangle$}
                                                           \put(0.1,1.1){$j_1$} \put(0.1,-0.10){$j_2$}
                                                           \put(-0.25,0.8){{\scriptsize X}} \put(-0.25,-0.4){${\scriptsize \hat{n}}$}
\vspace{0.5cm}
                            \\
                            \hspace{-0.2cm} \footnotesize{(b)} \hspace{0.3cm}
                                            \parbox{6.0cm}{
                                                         \epsfig{file=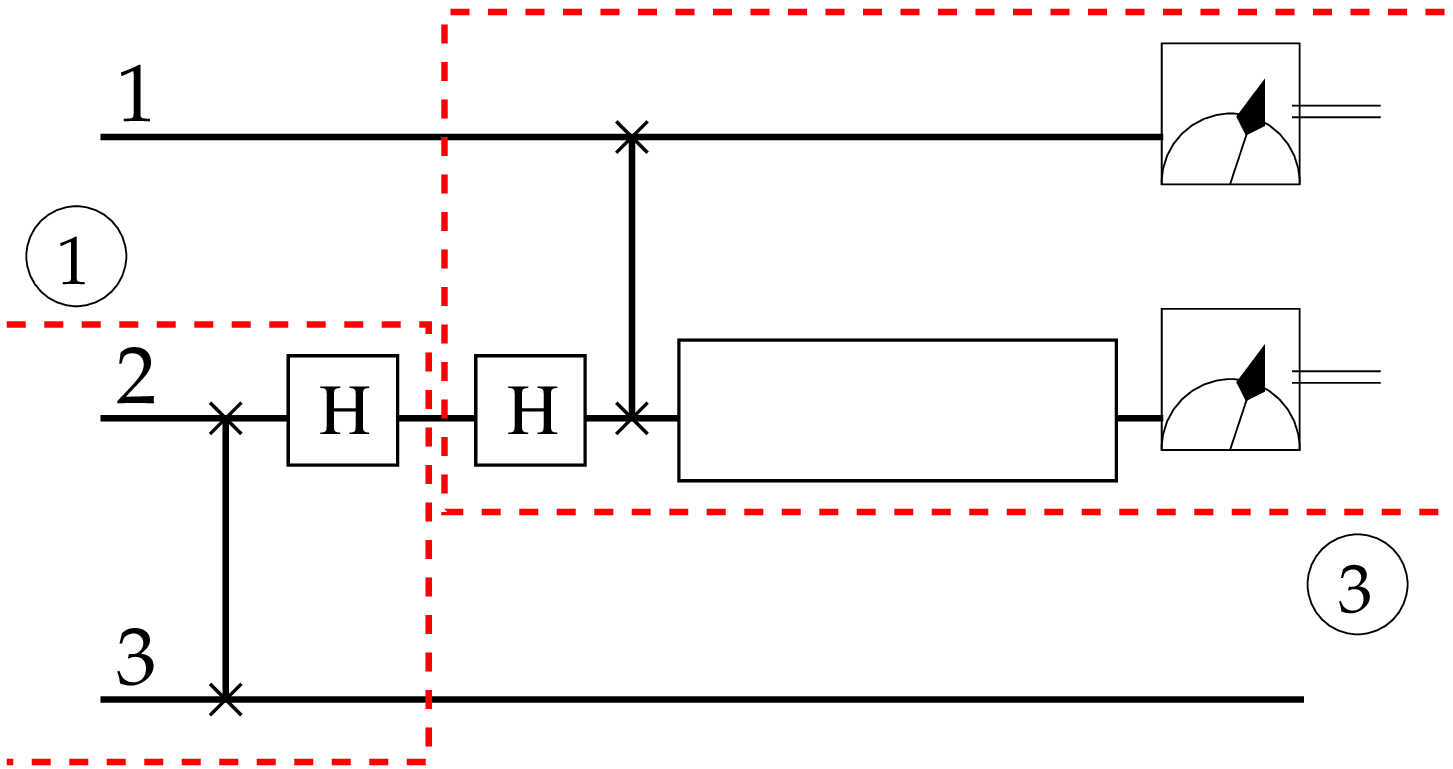,width=6.0cm}
                                                         } \put(-6.1,1.1){$|\Psi\rangle$} \put(-6.1,-0.1){$|+\rangle$} \put(-6.1,-1.25){$|+\rangle$}
                                                           \put(-0.1,1.15){$j_1$} \put(-0.1,0){$j_2$}
                                                           \put(-0.5,0.8){{\scriptsize X}} \put(-0.5,-0.35){{\scriptsize X}}
                                                           \put(-3.1,-0.1){$U_z((-1)^{j_1}\phi)$}
         \end{tabular}
\vspace{0.2cm}
\caption{\label{1wqc_xrot_network}
         \footnotesize{
                       In (a) we directly translate the measurement pattern shown in Fig.~\ref{cluster gates}a. In (b), inserting the identity in the form $H^2=I$ we identify box 1 as preparing a Bell state $|\Phi_0\rangle$ and box 3 as performing a generalized Bell measurement. Replacing the measurement along $\hat{n}$ with a rotation by $(-1)^{j_1}\phi$ about the $z$-axis followed by an measurement in the $X$-basis we complete the mapping from (a) to (b).
                      }
         }
     \end{center}
\end{figure}

In Fig.$\,$\ref{cluster gates}a we sketched the pattern that realizes a rotation by an angle $\phi$ about the $x$-axis in the 1WQC model. The first qubit is initially measured along the $x$-axis, and then the second is projected along the axis $\hat{n}=\{\cos((-1)^{j_1+1}\phi), \sin((-1)^{j_1+1}\phi),0\}$ on the equator of the Bloch sphere, conditioned on the outcome $j_1$ of the measurement of the first qubit. This is a typical example in the 1WQC of a measurement performed in a basis adapted according to previous measurement results. A circuit representation is given in Fig.$\,$\ref{1wqc_xrot_network}a.

\vspace{0.5cm}

\begin{figure}[b]
     \begin{center}
          \begin{tabular}{c}
                            \footnotesize{(a)} \hspace{0.1cm}
                                            \parbox{6.8cm}{
                                                         \epsfig{file=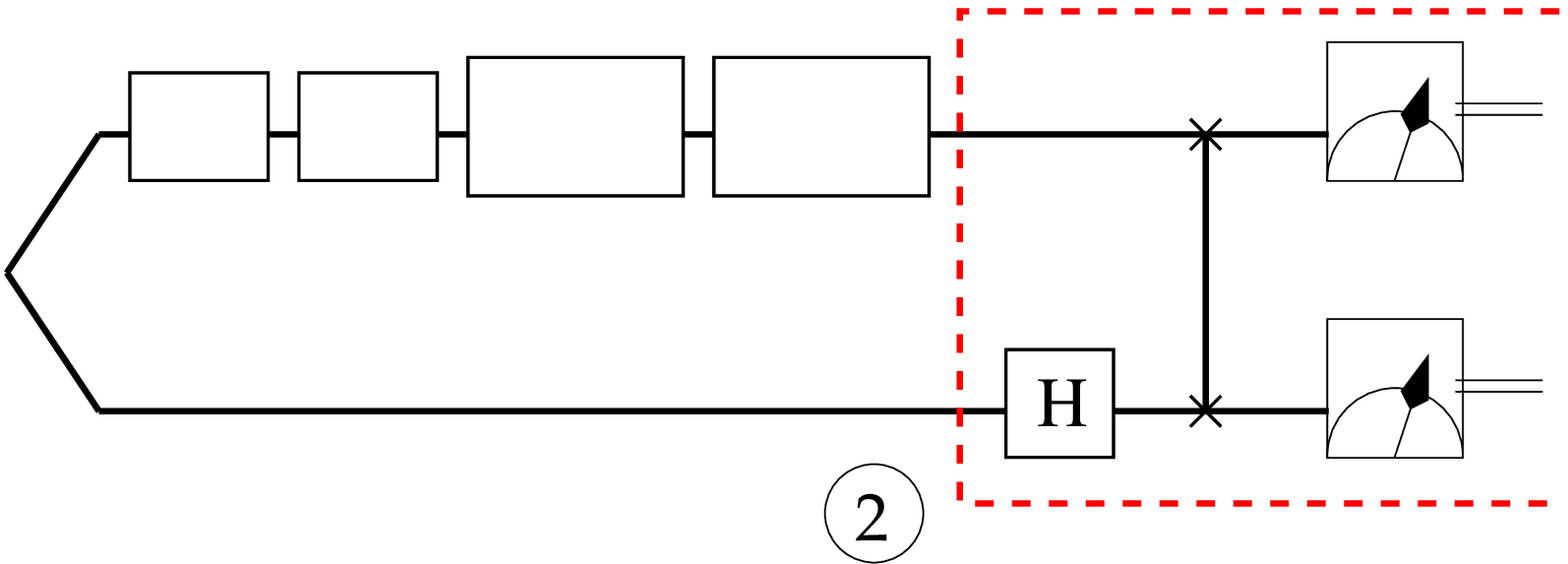,width=6.8cm}
                                                         } \put(0,0.8){$j_1$} \put(0,-0.4){$j_2$}
                                                           \put(-0.4,0.5){{\scriptsize X}} \put(-0.4,-0.75){{\scriptsize X}}
                                                           \put(-6.15,0.65){$Z^{j_1}$} \put(-5.45,0.65){$X^{j_2}$}
                                                           \put(-4.65,0.65){$U_x (\phi)$}
                                                           \put(-3.6,0.65){$U_x^\dagger (\phi)$}
\vspace{0.5cm}
                            \\
                            \footnotesize{(b)} \hspace{0.1cm}
                                            \parbox{7.0cm}{
                                                         \epsfig{file=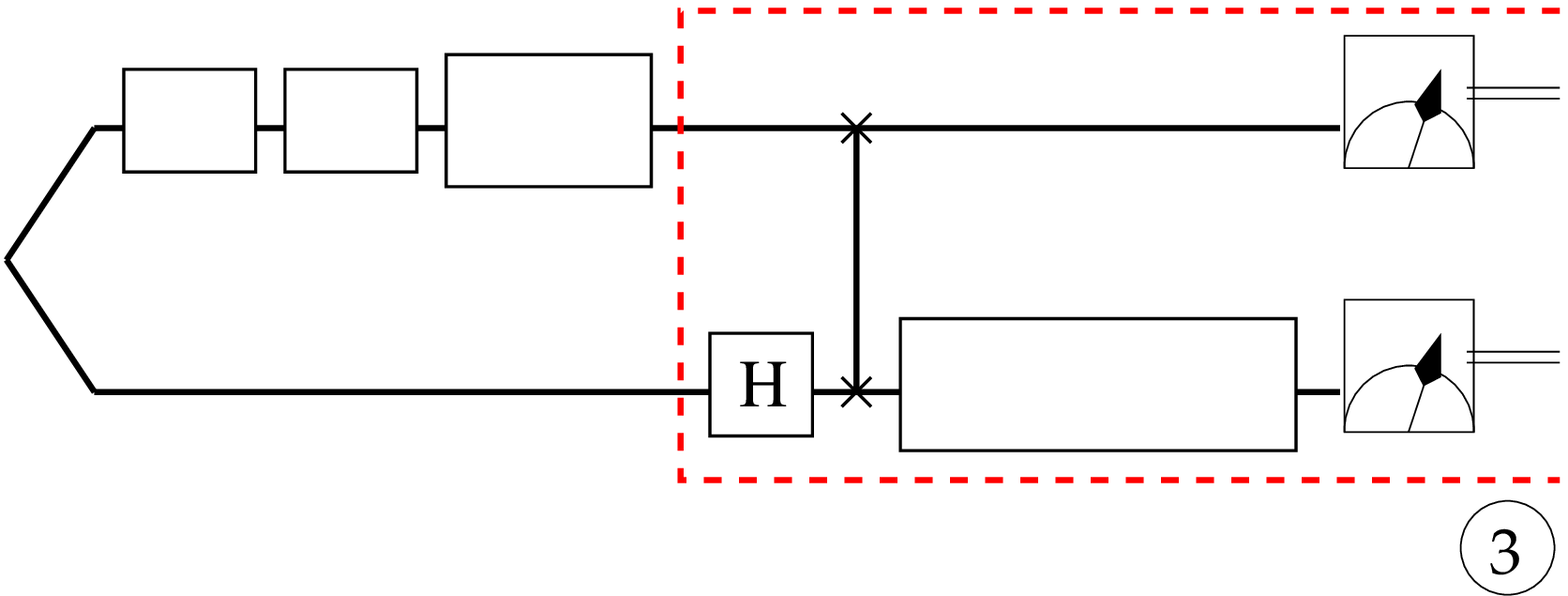,width=7.0cm}
                                                         } \put(0.1,0.9){$j_1$} \put(0.1,-0.2){$j_2$}
                                                           \put(-0.35,0.55){{\scriptsize X}} \put(-0.35,-0.6){{\scriptsize X}}
                                                           \put(-2.9,-0.39){$U_z((-1)^{j_1}\phi)$}
                                                           \put(-6.37,0.75){$Z^{j_1}$} \put(-5.67,0.75){$X^{j_2}$}
                                                           \put(-4.93,0.75){$U_x^\dagger (\phi)$}
         \end{tabular}
\caption{\label{xrot_identity}
         \footnotesize{
                      (a) Starting with the identity circuit that prepares the Bell state $|\Phi_j\rangle$ and then performs a Bell measurement, we insert the identity $U_x(\phi )U_x^\dagger (\phi )=I $. (b) Commute $U_x(\phi )$ to the left, then use the symmetry of $|\Phi_0\rangle$ in order to apply the rotation to the second qubit. Finally commute it to the right through the Hadamard to end up with box 3 of Fig.$\,$\ref{1wqc_xrot_network}b.
                       }
         }
     \end{center}
\end{figure}
\vspace{-0.5cm}
The transition from Fig.$\,$\ref{1wqc_xrot_network}a to Fig.$\,$\ref{1wqc_xrot_network}b is made again by inserting the identity $H^2=I$ and noticing that the measurement along the $\hat{n}$-axis can be replaced by the rotation $U_z((-1)^{j_1}\phi)$ followed by a measurement in the $X$-basis. In order to interpret the generalized measurement performed by box 3, we would like to translate it into a measurement in the basis $\{(U^\dagger \otimes I)|\Phi_j\rangle\}_j$, where $U$ is the unitary to be applied to the input state, to match the TQC operation (Fig.~\ref{teleportation}b). We first note that, starting with the Bell state $|\Phi_0\rangle$, the other three Bell states can be created by applying $Z^{j_1}X^{j_2}$ on the first qubit
\[
   |\Phi_j\rangle=(Z^{j_1}X^{j_2}\otimes I) |\Phi_0\rangle ,\hspace{0.2cm} j=(j_1,j_1\oplus j_2).
\]
The identity of the Bell state can subsequently be revealed by the Bell measurement performed by the circuit of Fig.~\ref{identities}b or equivalently box 2 of Fig.$\,$\ref{1wqc_tel_network}b. Before attaching the Bell state $|\Phi_j\rangle$ to the Bell measurement, we insert the identity in the form $U_x(\phi )U_x^\dagger (\phi )=I $ to the first qubit as shown in Fig.$\,$\ref{xrot_identity}a. Commuting $U_x(\phi )$ to the left through $Z^{j_1}$ it becomes $U_x((-1)^{j_1}\phi )$, which can equivalently be applied to the second qubit. Here we made use of the useful symmetry property of the Bell state $|\Phi_0\rangle$ \pagebreak
\[
     (U\otimes I)|\Phi_0\rangle = (I\otimes U^T)|\Phi_0\rangle ,
\]
which is true for any operator $U$. Then $U_x((-1)^{j_1}\phi )$ can be commuted to the right of the Hadamard giving $U_z((-1)^{j_1}\phi )$, which commutes with the controlled-phase. Shifting $U_x^\dagger (\phi )$ on the first qubit to the left into the Bell state preparation and with $U_z((-1)^{j_1}\phi )$ now inside the Bell measurement on the second qubit, we finally obtain an interpretation of box 3 as shown by the identity circuit of Fig.$\,$\ref{xrot_identity}b: the state $(U_x^\dagger (\phi)\otimes I)|\Phi_j\rangle$ is first prepared and then measured.

Hence box 3 in Fig.$\,$\ref{1wqc_xrot_network}b indeed performes a measurement in the basis $\{(U_x^\dagger(\phi ) \otimes I)|\Phi_j\rangle\}_j$. This completes the mapping for $U_x(\phi )$ from the 1WQC to the TQC. Again, the reverse mapping is easily obtained by tracing all steps backwards.

\vspace{0.0cm}
\subsection{Rotation about the $z$-axis}
\begin{figure}[htb]
     \begin{center}
          \begin{tabular}{c}
                            \hspace{-1cm} {\footnotesize (a)} \hspace{0.7cm}
                                            \parbox{4.2cm}{
                                                         \epsfig{file=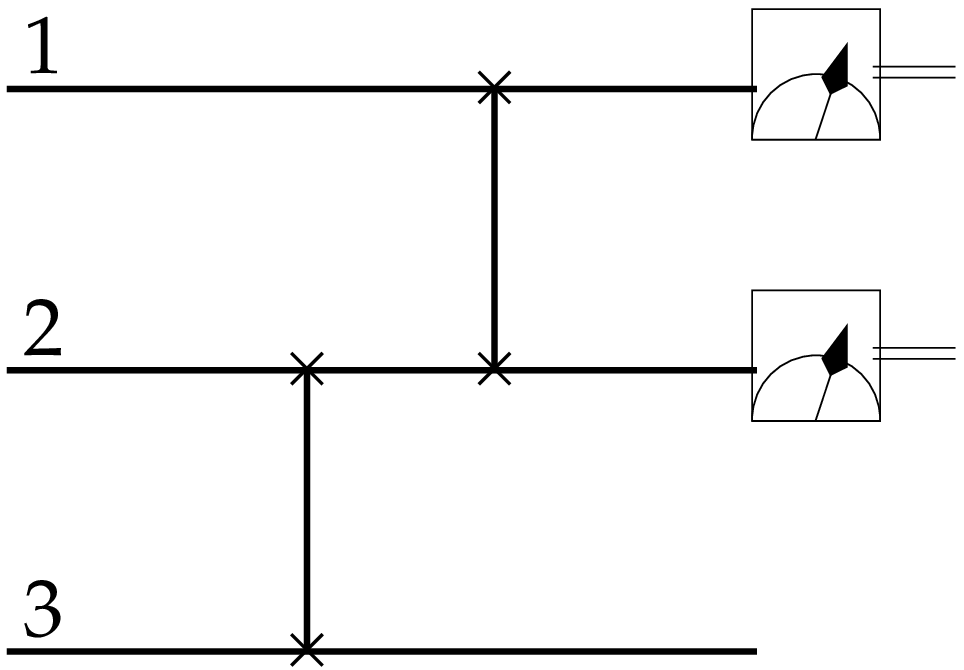,width=4.2cm}
                                                         } \put(-4.65,1.1){$|\Psi\rangle$} \put(-4.65,-0.1){$|+\rangle$} \put(-4.65,-1.3){$|+\rangle$}
                                                           \put(0.1,1.1){$j_1$} \put(0.1,-0.10){$j_2$}
                                                           \put(-0.25,0.8){${\scriptsize \hat{k}}$} \put(-0.25,-0.4){{\scriptsize X}}
\vspace{0.5cm}
                            \\
                            \hspace{-0.4cm} {\footnotesize (b)} \hspace{0.5cm}
                                            \parbox{5.6cm}{
                                                         \epsfig{file=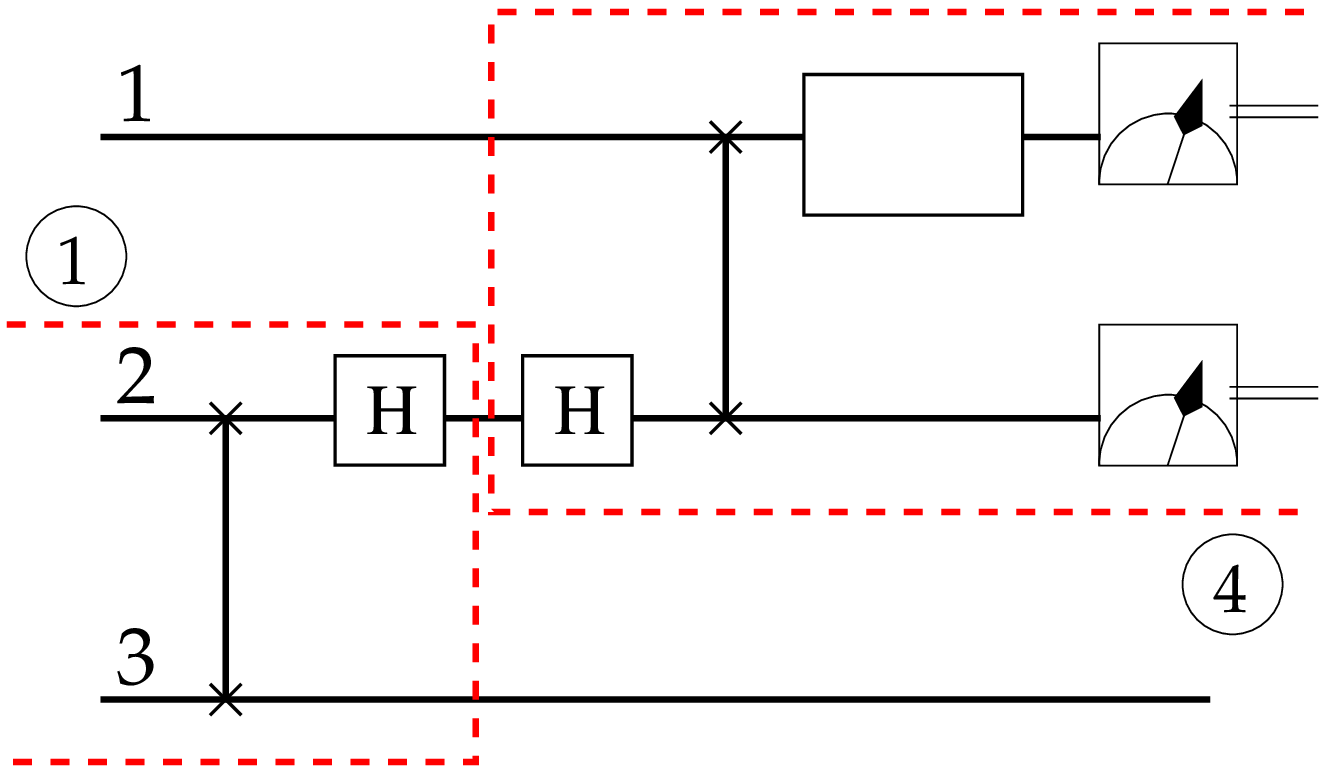,width=5.6cm}
                                                         } \put(-5.7,1.15){$|\Psi\rangle$} \put(-5.7,-0.1){$|+\rangle$} \put(-5.7,-1.2){$|+\rangle$}
                                                           \put(0.1,1.2){$j_1$} \put(0.1,0){$j_2$}
                                                           \put(-2.1,1.0){$U_z(\theta )$}
                                                           \put(-0.25,0.9){{\scriptsize X}} \put(-0.25,-0.4){{\scriptsize X}}
         \end{tabular}
\caption{\label{1wqc_zrot_network}
         \footnotesize{
                       In (a) we translate the 1WQC pattern shown in Fig.~\ref{cluster gates}b. Replacing the measurement along $\hat{k}$ with a rotation by $\theta$ around the $z$-axis followed by an X-measurement we obtain the mapping from (a) to (b).
                      }
         }
     \end{center}
\end{figure}

A rotation by an angle $\theta$ about the $z$-axis is realized in the 1WQC model by the qubit pattern of Fig.$\,$\ref{cluster gates}b. The first qubit is projected along the direction $\hat{k}=\{\cos(-\theta), \sin(-\theta),0\}$ in the Bloch sphere, whereas the second is projected along the $x$-axis. This time no special time ordering is necessary, so both measurements can be done simultaneously. An equivalent circuit representation is given by Fig.$\,$\ref{1wqc_zrot_network}a. The measurement along the $\hat{k}$-axis can be replaced by the rotation $U_z((-1)^{j_1}\phi)$ followed by a measurement in the $X$-basis. Inserting once more the identity $H^2=I$, we obtain the circuit of Fig.$\,$\ref{1wqc_zrot_network}b.

\vspace{0.5cm}

\begin{figure}[tb]
     \begin{center}
              \epsfig{file=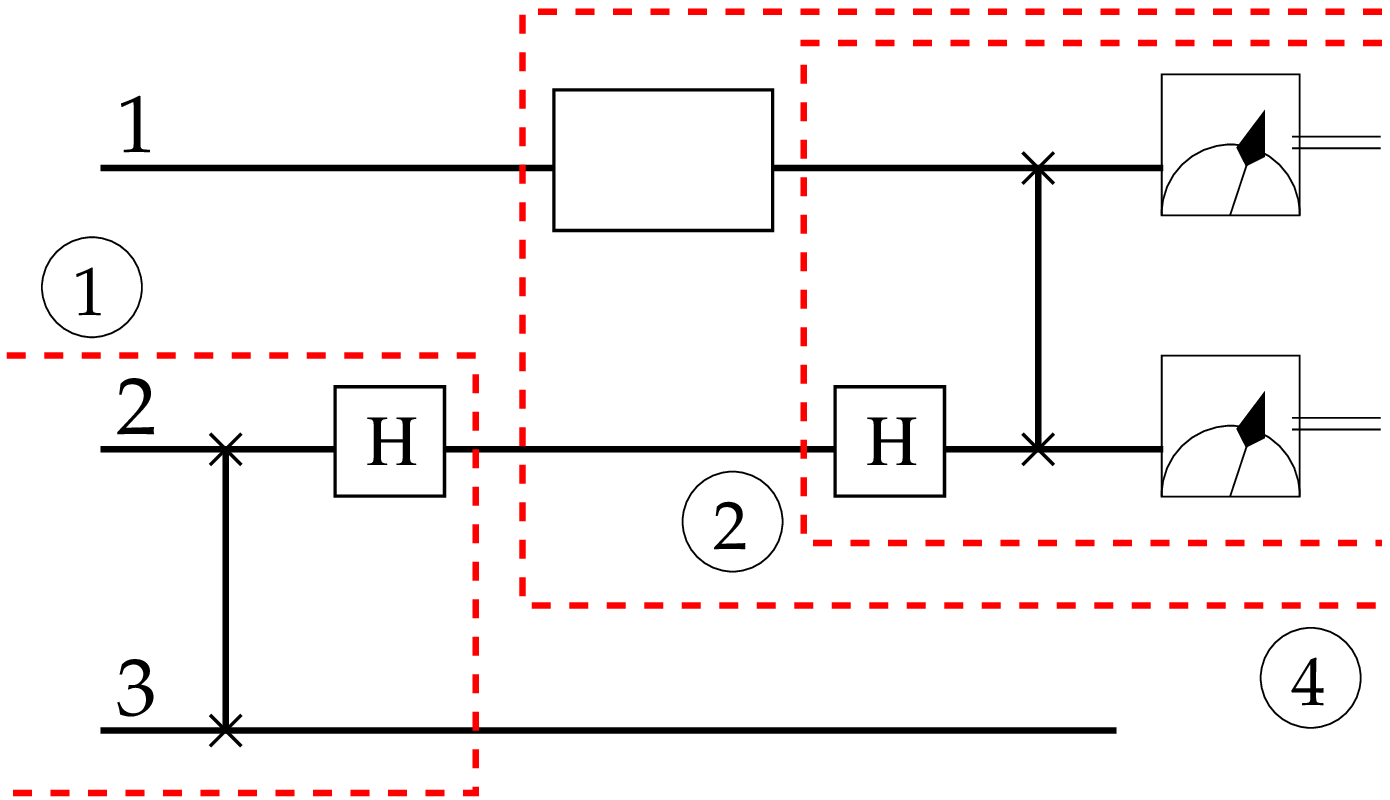,width=5.7cm}     \put(-5.83,2.5){$|\Psi\rangle$} \put(-5.83,1.35){$|+\rangle$} \put(-5.83,0.23){$|+\rangle$}
                                                           \put(0.1,2.59){$j_1$} \put(0.1,1.44){$j_2$}
                                                           \put(-3.35,2.5){$U_z(\theta )$}
                                                           \put(-0.27,2.25){{\scriptsize X}} \put(-0.27,1.15){{\scriptsize X}}

\caption{\label{1wqc_zrot_equiv}
         \footnotesize{Since box 2 realizes a measurement in the Bell basis $\{|\Phi_j\rangle\}_j$,
box 4 is a measurement in the basis $\{(U_z^\dagger(\theta ) \otimes I)|\Phi_j\rangle\}_j$ 
according to Fig.~\ref{teleportation}b.
                      }
         }  
     \end{center}
\end{figure}

In order to interpret the measurement performed by box 4 in Fig.$\,$\ref{1wqc_zrot_network}b, we note that $U_z(\theta)$ commutes with the controlled-phase gate and therefore can be taken out of the measurement box as sketched in Fig.~\ref{1wqc_zrot_equiv}. Hence, box 4 can be thought of as a rotation $U_z(\phi)$ applied to the input, followed by a Bell measurement. This is precisely equivalent to a measurement in the basis $\{(U_z^\dagger(\theta ) \otimes I)|\Phi_j\rangle\}_j$.

This completes the mapping for $U_z(\theta)$ from the 1WQC to the TLC, which again can equally well be traced in the opposite direction.

\vspace{0.3cm}

\vspace{0.0cm}
\subsection{The controlled-NOT}
\begin{figure}[htb]
              \epsfig{file=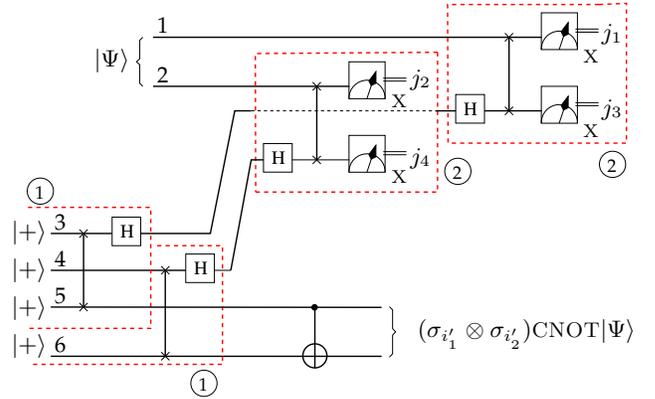,width=8cm} \put(-2.8,0.8){$(\sigma_{i_1'}\otimes \sigma_{i_2'})\text{{\footnotesize \sc CNOT}}|\Psi\rangle$}
                                                     \put(-7.1,4.4){$|\Psi\rangle$}
                                                     \put(-8.2,2.1){$|+\rangle$} \put(-8.2,1.6){$|+\rangle$} \put(-8.2,1.1){$|+\rangle$} \put(-8.2,0.6){$|+\rangle$}
                                                     \put(-0.35,4.75){$j_1$} \put(-0.35,3.75){$j_3$} \put(-2.9,4.15){$j_2$} \put(-2.9,3.1){$j_4$}
                                                     \put(-0.6,4.45){{\scriptsize X}} \put(-0.6,3.45){{\scriptsize X}} \put(-3.15,3.85){{\scriptsize X}} \put(-3.15,2.85){{\scriptsize X}}
                                                     
\caption{\label{1wqc_cnot_network}
         \footnotesize{
The {\sc cnot} of Fig.~\ref{CNOT}, substituting boxes 1 and 2 from Fig.~\ref{1wqc_tel_network}b. The Pauli corrections $\sigma_{i_1'}\otimes \sigma_{i_2'}$ are given in terms of the measurement outcomes by the commutation relations $\text{{\sc cnot}}(\sigma_{i_1}\otimes \sigma_{i_2})=(\sigma_{i_1'}\otimes \sigma_{i_2'})\text{{\sc cnot}}$, where $i_1=(j_1,j_1\oplus j_3)$, $i_2=(j_2,j_2\oplus j_4)$.  
                      }
         }  
\end{figure}

To construct the mapping between the two models for the {\sc cnot} gate, it is easiest to begin with the teleportation-based circuit given in Fig.$\,$\ref{CNOT}. We then replace the ancillary Bell states $|\Phi_0\rangle$ and the Bell measurements by boxes 1 and 2 of Fig.$\,$\ref{1wqc_tel_network}b respectively. Thus we directly obtain the equivalent circuit of Fig.$\,$\ref{1wqc_cnot_network}.

\vspace{0.2cm}
\vspace{0.2cm}

\begin{figure}[htb]
    \begin{center}
      \begin{tabular}{cc}
                         \footnotesize{(a)} \hspace{0.4cm} & \footnotesize{(b)} \vspace{0.2cm}
                         \\
                         \epsfig{file=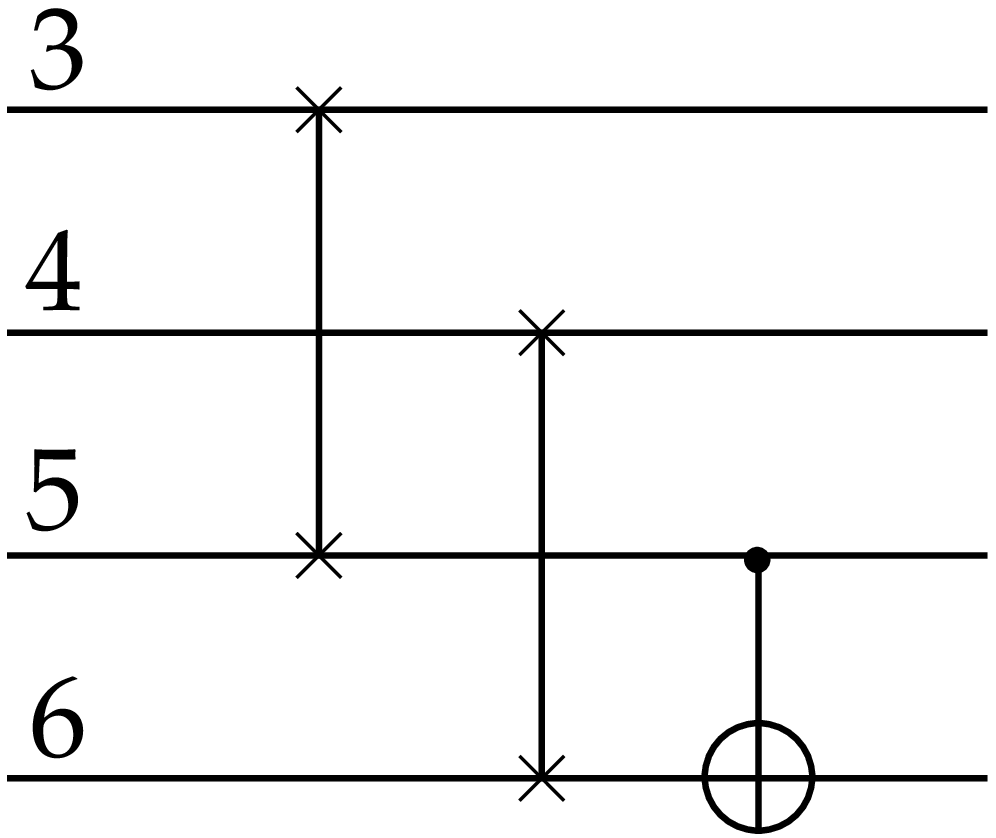,width=2.8cm} \put(-3.4,2.0){$|+\rangle$} \put(-3.4,1.35){$|+\rangle$} \put(-3.4,0.75){$|+\rangle$} \put(-3.4,0.15){$|+\rangle$}
                                                                  \hspace{1cm} &
                         \epsfig{file=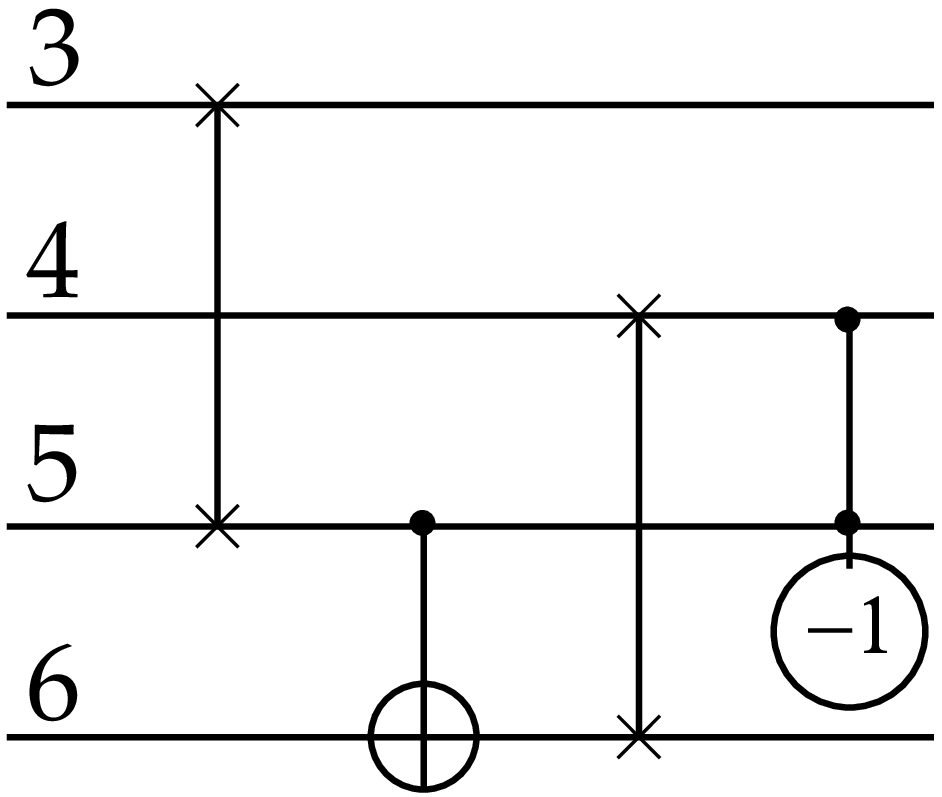,width=2.8cm} \put(-3.4,2.0){$|+\rangle$} \put(-3.4,1.35){$|+\rangle$} \put(-3.4,0.75){$|+\rangle$} \put(-3.4,0.15){$|+\rangle$}
                         \\ & \vspace{0.3cm} \\
                         \footnotesize{(c)} \hspace{0.4cm} & \footnotesize{(d)} \vspace{0.2cm}
                         \\
                         \epsfig{file=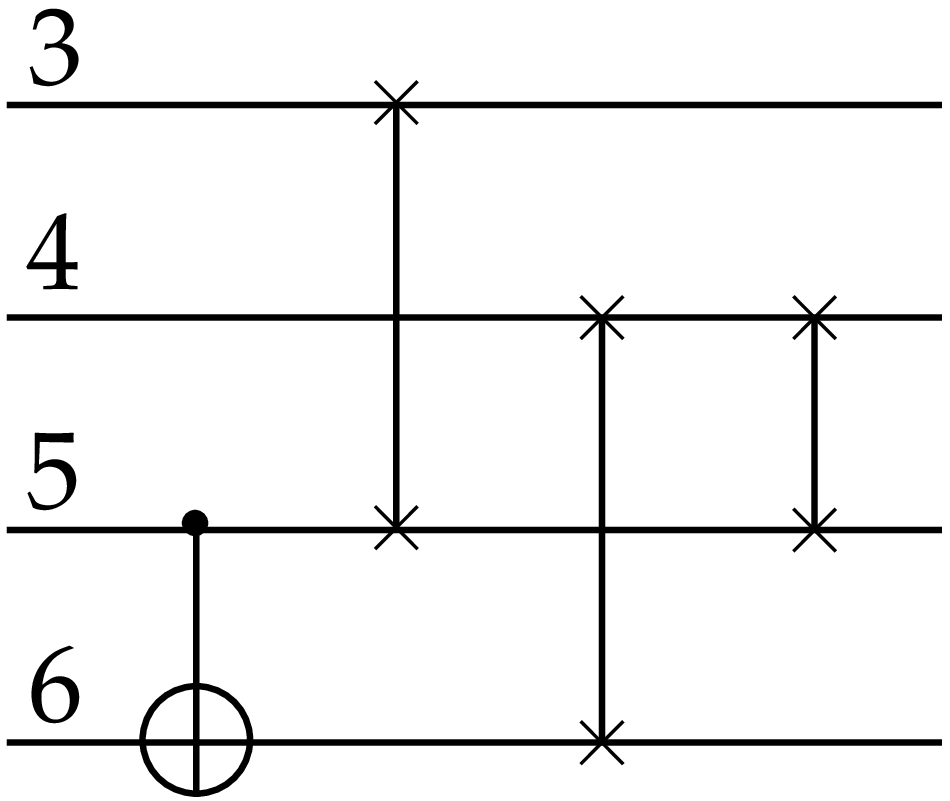,width=2.8cm} \put(-3.4,2.0){$|+\rangle$} \put(-3.4,1.35){$|+\rangle$} \put(-3.4,0.75){$|+\rangle$} \put(-3.4,0.15){$|+\rangle$}
                                                                  \hspace{0.8cm} &
                         \epsfig{file=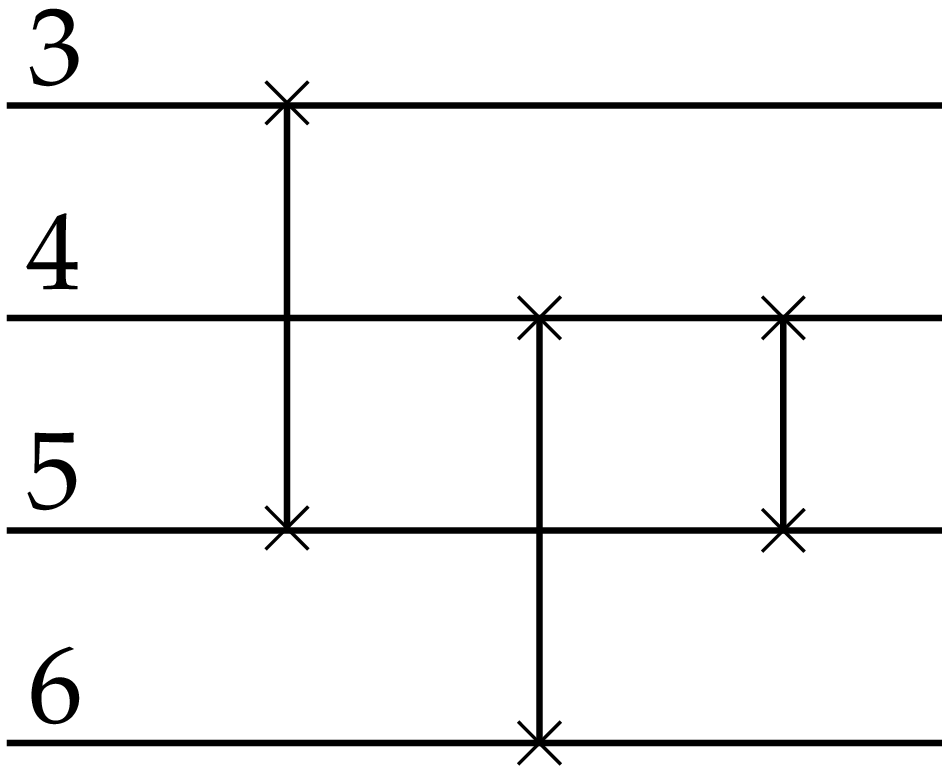,width=2.92cm} \put(-3.5,1.9){$|+\rangle$} \put(-3.5,1.25){$|+\rangle$} \put(-3.5,0.65){$|+\rangle$} \put(-3.5,0.05){$|+\rangle$}
      \end{tabular} \vspace{0.2cm}
\caption{\label{CNOT_explain}
         \footnotesize{
                       Successively commuting the {\sc cnot}$_{5\rightarrow 6}$ to the left we obtain an equivalent circuit that uses only controlled-phase gates. In (c) note that {\sc cnot}$|+\rangle |+\rangle=|+\rangle |+\rangle$, hence (d).
                       }
         }
    \end{center}
\end{figure}

We immediately note that the Hadamards cancel each other out on the third and fourth qubit, so that the circuit contains only controlled-phase gates and the {\sc cnot}$_{5\rightarrow 6}$. Since in the 1WQC the only unitary that acts between the cluster qubits is the controlled-phase, $\Lambda(Z)$, we seek a way to replace the {\sc cnot}$_{5\rightarrow 6}$ with a controlled-phase gate applied between some other pair of qubits. Inspecting the lower part of the circuit, we observe that the {\sc cnot} can be commuted through the $\Lambda(Z)_{(4,6)}$ leaving behind a factor that multiplies the state with $-1$ whenever both the fourth and the fifth qubit are in the state $|1\rangle$. This is exactly how a $\Lambda(Z)_{(4,5)}$ gate would act. Fig.$\,$\ref{CNOT_explain} summarizes the argument.

\vspace{0.2cm}

\begin{figure}[htb]
      \begin{center}
              \epsfig{file=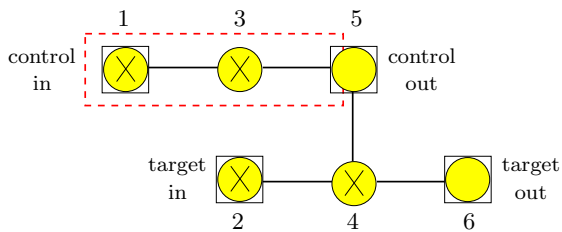,width=5.45cm}
                                                        \put(-4.7,0.3){\parbox{1cm}{\footnotesize{target in}}} \put(0,0.3){\parbox{1cm}{\footnotesize{target out}}}
                                                        \put(-6.5,1.75){\parbox{1cm}{\footnotesize{control in}}} \put(-1.45,1.75){\parbox{1cm}{\footnotesize{control out}}}
                                                        \put(-5.0,2.4){1}  \put(-3.45,2.4){3} \put(-1.9,2.4){5}
                                                        \put(-3.48,-0.3){2}  \put(-1.95,-0.3){4} \put(-0.4,-0.3){6} 
\vspace{0.2cm}
\caption{\label{1wqc_cnot_final}
         \footnotesize{
The realization of {\sc cnot} in the 1WQC. The control qubit is teleported from 1 to 5 by the three-qubit wire pattern of Fig.$\,$\ref{1wqc_tel}. The rest of the circuit is exactly the pattern for {\sc cnot} in Fig.~\ref{cluster gates}c.
                      }
         }  
      \end{center}
\end{figure}

Therefore the circuit of Fig.$\,$\ref{1wqc_cnot_network} can be transformed into an equivalent circuit that makes use of qubits prepared in the $|+\rangle$ state, controlled-phase gates and one-qubit measurements in the $X$-basis. Hence it can directly be mapped into the pattern of the 1WQC drawn in Fig.$\,$\ref{1wqc_cnot_final}.

\vspace{0.2cm}

\begin{figure}[htb]
      \begin{center}
              \epsfig{file=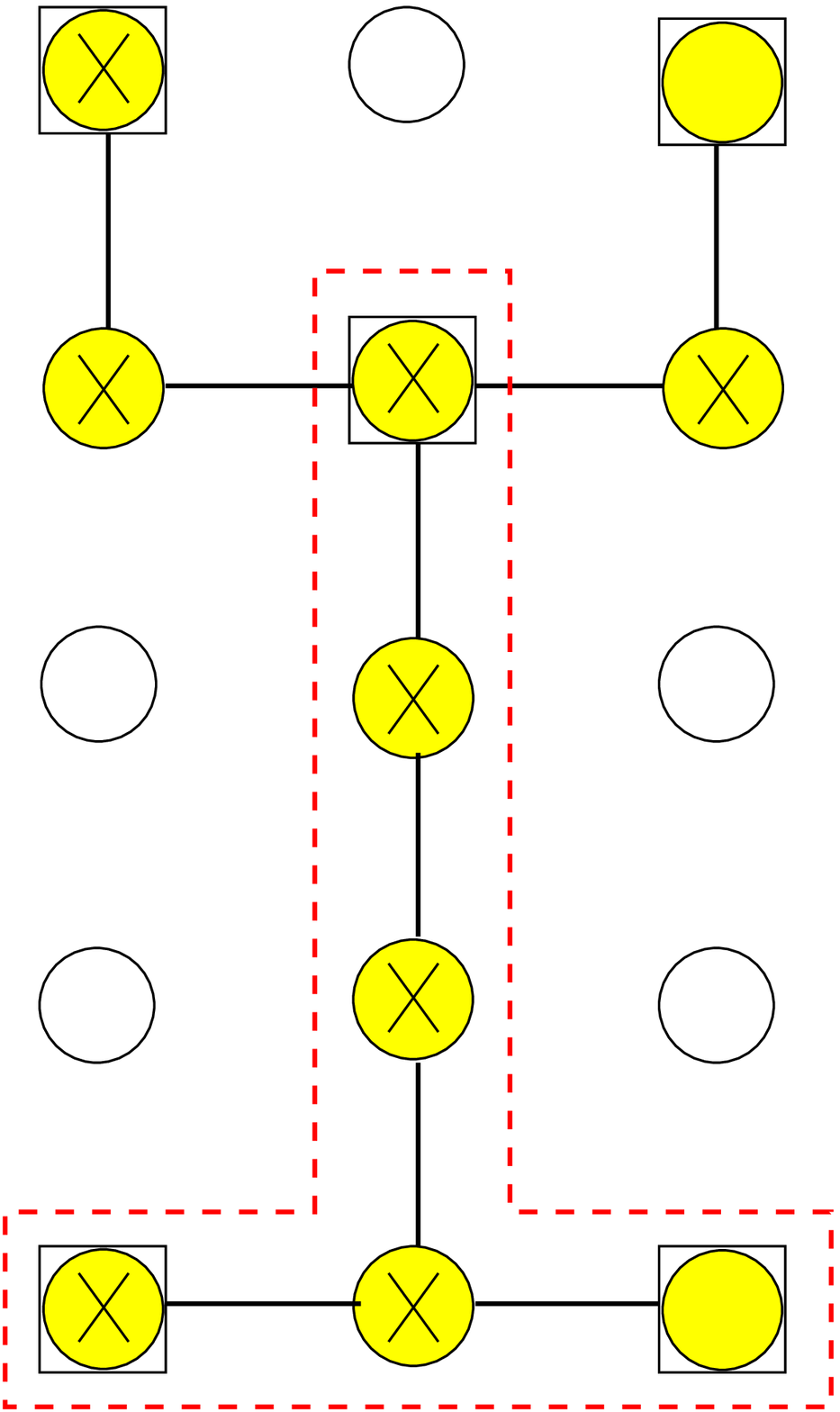,width=3.5cm} \put(-4.5,0.3){\parbox{1cm}{\footnotesize{target in}}} \put(0,0.3){\parbox{1cm}{\footnotesize{target out}}}
                                                     \put(-4.5,5.5){\parbox{1cm}{\footnotesize{control in}}} \put(-0.1,5.5){\parbox{1cm}{\footnotesize{control out}}}
                                                     \put(-3.6,5.0){1}  \put(-3.6,3.9){2} \put(-0.95,3.9){4} \put(-0.95,5.0){5}
                                                     \put(-2.5,3.9){3}  \put(-2.5,2.9){6} \put(-2.5,1.65){7}
                                                     \put(-3.2,-0.3){8}  \put(-1.85,-0.3){9} \put(-0.6,-0.3){10} 
\vspace{0.2cm}
\caption{\label{1wqc_cnot_square}
         \footnotesize{
                       The control is teleported from qubit 1 to 3 before the {\sc cnot}, marked by the dashed line, is applied. The control is then teleported again from site 3 to 5. The pattern from qubit 3 to 9 realizes a controlled-phase between those two qubits, as is explained in the following section. Empty circles indicate qubits removed from the cluster after being measured in the $Z$-basis. 
                      }
         }  
      \end{center}
\end{figure}

We note that the resulting 1WQC construction for the {\sc cnot} first teleports the control state from the first to the fifth qubit before the {\sc cnot} is applied. In fact, using extra teleportation steps for the control qubit is necessary for such a {\sc cnot} construction in the 1WQC \cite{raussen01}, since it allows the explicit separation of the input from the output qubits and therefore enables the identification of the input (output) qubits of {\sc cnot} with the output (input) qubits of the gate preceeding (following) it in the computation. This is particularly important when computation is performed on a two-dimensional cluster, where the geometry of the lattice imposes restrictions on how patterns, corresponding to different gates, can be composed in a sequence. An example of how the {\sc cnot} can be combined with the quantum wire of Fig.$\,$\ref{1wqc_tel} to give a square-shaped pattern in a two-dimensional cluster is shown in Fig.$\,$\ref{1wqc_cnot_square}.

\vspace{0.2cm}

Thus, we have completed the mapping from the TQC to the 1WQC for the {\sc cnot}, which again can be worked backwards as long as the extra teleportation step is included.

At this point, we have reached our goal of demonstrating a systematic mapping between the two models for the universal set of one-qubit rotations and {\sc cnot}. The two models are therefore shown to make use of interchangeable primitives in order to implement computation by measurement alone. Depending on which model we consider as more fundamental, or conceptually easier to understand, we can group in boxes operations performed in the 1WQC to obtain primitive operations of the TQC (Bell pairs and complete two-qubit measurements), or alternatively we can decompose the primitive operations of the TQC into the primitives of the 1WQC (controlled-phase gates and one-qubit measurements). 

\vspace{0.0cm}
\section{A simple remote-$\mathbf{\Lambda(Z)}$ circuit and its applications}

In this section, we describe a simple method of implementing the remote-$\Lambda(Z)$ gate that applies to both the 1WQC and the TQC models. As will subsequently be shown, this construction leads to significant simplifications in the basic operations in both models and can also be applied for the preparation of arbitrary graph states using two-qubit measurements alone. The key observation that leads to these simplifications is that given the ability to perform single qubit gates, both the {\sc cnot} and the $\Lambda(Z)$ can equally well be used to complete universality.

\begin{figure}[htb]
      \begin{center}
              \epsfig{file=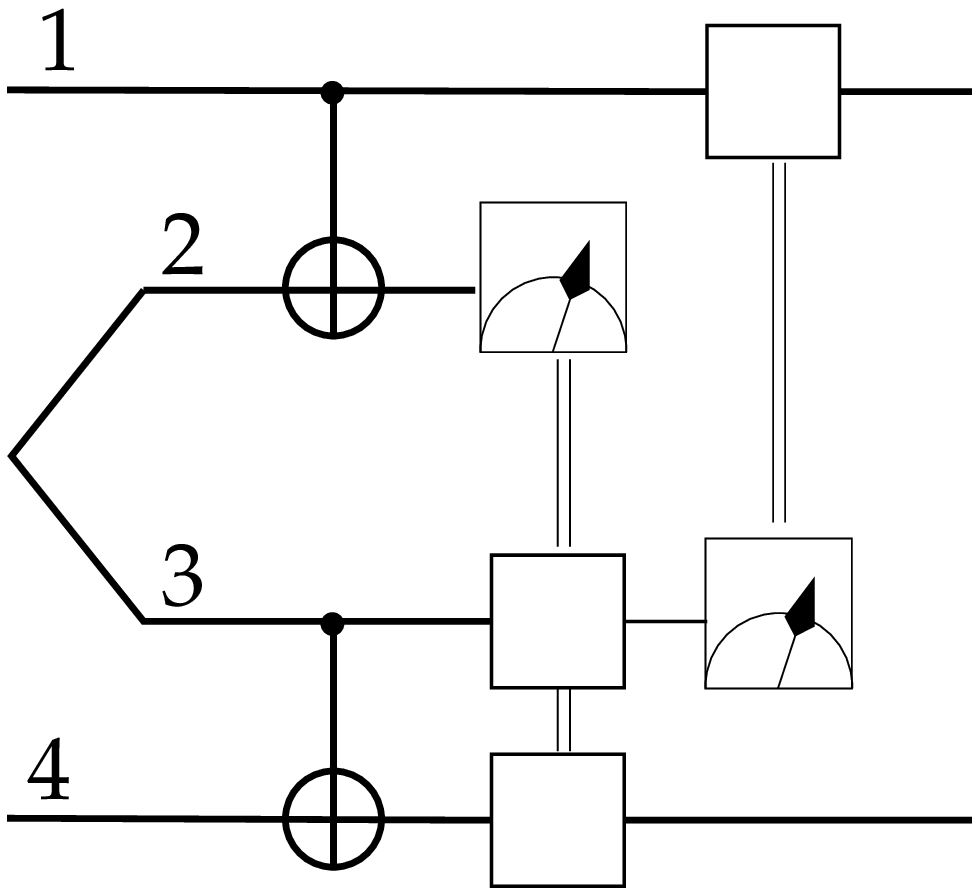,width=4cm} \put(-5.05,3.2){\footnotesize{control}} \put(-4.95,0.25){\footnotesize{target}}
                                                      \put(-1.87,1.0){$X$} \put(-1.87,0.17){$X$} \put(-0.95,3.15){$Z$}
                                                      \put(-1.4,2.15){{\scriptsize Z}} \put(-0.45,0.75){{\scriptsize X}}                                         
\vspace{0.1cm}
\caption{\label{remote_cnot}
         \footnotesize{
The remote-{\sc cnot}: We perform {\sc cnot} gates from the control qubit to one half of a Bell pair and also from the other Bell pair qubit to the target qubit. By subsequently measuring the Bell pair qubits, the {\sc cnot}$_{1\rightarrow 4}$ gate is realized (up to Pauli corrections) without qubits 1 and 4 having directly interacted. 
                      }
         }  
      \end{center}
\end{figure}

In the TQC, we recall that the {\sc cnot} is realized by the circuit of Fig.$\,$\ref{CNOT}, using a special ancillary state, $|a_{\text{{\sc cnot}}}\rangle$, and two Bell measurements. Considering that the preparation of $|a_{\text{{\sc cnot}}}\rangle$ requires three two-qubit measurements \cite{leung01}, the total number of two-qubit measurements needed for this procedure is five. Here we will employ two-qubit incomplete measurements in order to implement the controlled-phase gate in a significantly more resource-effective way. Our starting point will be the remote-{\sc cnot} gate that has appeared in the literature \cite{gottesman01}, shown in Fig.$\,$\ref{remote_cnot}. 

\vspace{0.2cm}

\begin{figure}[tb]
    \begin{center}
      \begin{tabular}{cccc}
                       \multicolumn{4}{c}{ \hspace{0.8cm} \footnotesize{(a)} \hspace{0.8cm} 
                                                                                 \parbox{5cm}{
                                                                                 \epsfig{file=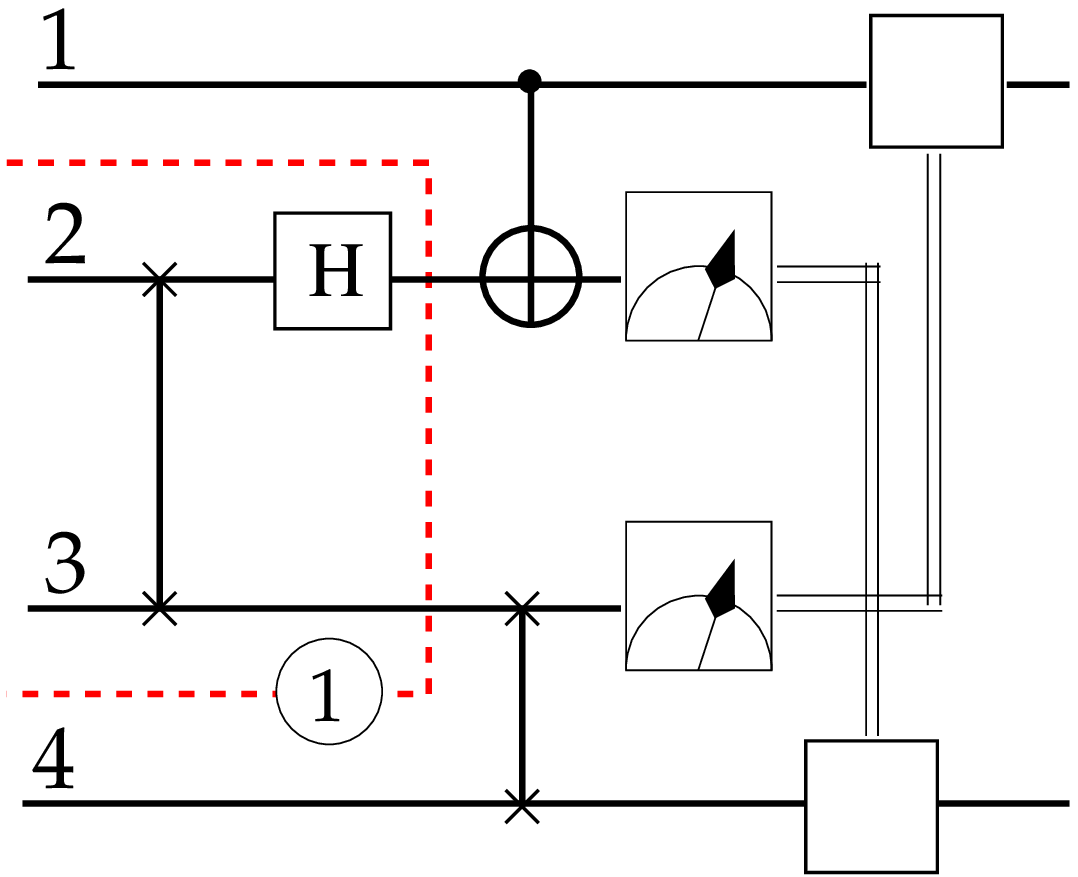,width=4.55cm}
                                                                                               }                                          
                                                                                } 
                                                                                 \put(-5.7,1.5){\footnotesize{control}} \put(-5.65,-1.5){\footnotesize{target}}
                                                                                 \put(-1.5,0.45){{\scriptsize Z}} \put(-1.5,-1.0){{\scriptsize X}}
                                                                                 \put(-1.28,-1.58){$Z$} \put(-1.0,1.5){$Z$}
                                                                                 \put(-5.3,0.7){$|+\rangle$} \put(-5.3,-0.7){$|+\rangle$}
                         \\ & & & \vspace{0.3cm} \\
                        \hspace{-0.2cm} \footnotesize{(b)} \hspace{0.5cm} & \parbox{4.0cm}{
                                                                            \epsfig{file=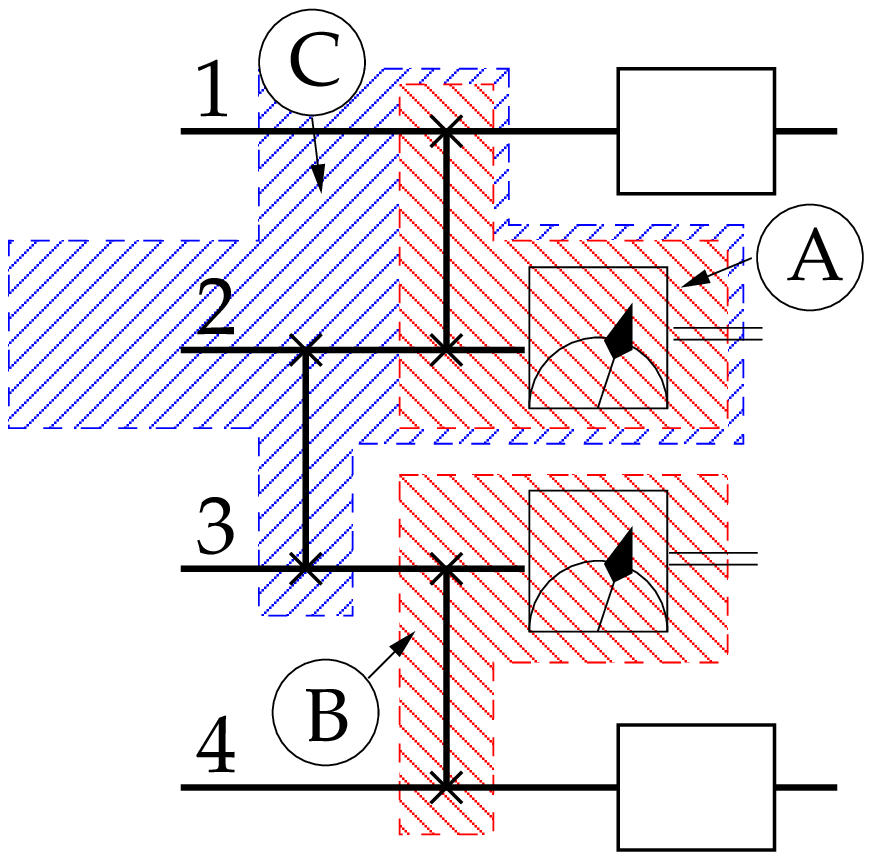,width=4.0cm}
                                                                           }
                                                                                   \put(-4.2,1.4){\footnotesize{control}} \put(-4.05,-1.65){\footnotesize{target}}
                                                                                   \put(-0.9,0.15){{\scriptsize X}} \put(-0.9,-0.95){{\scriptsize X}}
                                                                                   \put(-0.3,0.34){$j_2$} \put(-0.3,-0.6){$j_3$}
                                                                                   \put(-3.8,0.4){$|+\rangle$} \put(-3.8,-0.6){$|+\rangle$}
                                                                                   \put(-1.0,-1.73){$Z^{j_2}$} \put(-1.0,1.32){$Z^{j_3}$}
                                                                                   \hspace{0.4cm} &
                          \footnotesize{(c)} \hspace{0.5cm} & \parbox{0.55cm}{
                                                               \epsfig{file=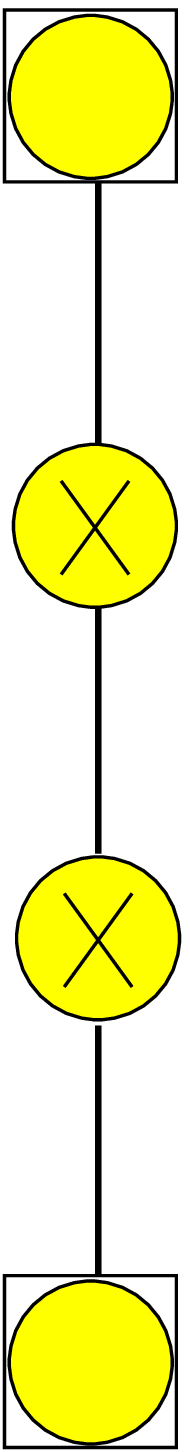,width=0.55cm}
                                                              }
                                                                                   \put(-0.75,1.85){1} \put(-0.75,0.58){2} \put(-0.75,-0.7){3} \put(-0.75,-1.95){4}
      \end{tabular} \vspace{0.2cm}
\caption{\label{remote_C-phase}
         \footnotesize{
                       (a) The remote-$\Lambda(Z)$, (b) an equivalent circuit that uses only controlled-phase gates and one-qubit measurements and (c) its direct translation in the 1WQC which realizes $\Lambda(Z)_{(1,4)}$ after qubits are measured in the $X$-basis.
                       }
         }
    \end{center}
\end{figure}

Starting from Fig.$\,$\ref{remote_cnot}, one can obtain a method of performing the remote-$\Lambda(Z)$ as shown in Fig.$\,$\ref{remote_C-phase}a: conjugating the fourth qubit with Hadamards, the {\sc cnot}$_{3\rightarrow 4}$ is replaced by $\Lambda(Z)_{(3,4)}$, according to the identity of Fig.$\,$\ref{identities}c, with the correction being switched from $\sigma_x$ to $\sigma_z$. Furthermore, in Fig.$\,$\ref{remote_C-phase}a we have used box 1 of Fig.$\,$\ref{1wqc_tel_network}b that provides an equivalent way to prepare the Bell state $|\Phi_0\rangle$. Surprisingly enough, the resulting circuit can be transformed to consist only of controlled-phase gates and one-qubit measurements in the $X$-basis. Indeed, using the identity of Fig.$\,$\ref{identities}c once more, the {\sc cnot}$_{1\rightarrow 2}$ can be replaced by $\Lambda(Z)_{(1,2)}$ with the Hadamard on its left side being cancelled out and the one on its right turning the measurement of the second qubit into a measurement in the $X$-basis. Thus we obtain the circuit of Fig.$\,$\ref{remote_C-phase}b, which can be directly translated into a four qubit pattern that realizes the controlled-phase gate between the first and fourth qubit in the 1WQC as shown in Fig.$\,$\ref{remote_C-phase}c.

\vspace{0.2cm}

\vspace{0.0cm}
\subsection{Simplifying the 1WQC model}

\begin{figure}[htb]
     \begin{center}
          \begin{tabular}{c}
                            \footnotesize{(a)} \vspace{0.2cm} \\
                            \epsfig{file=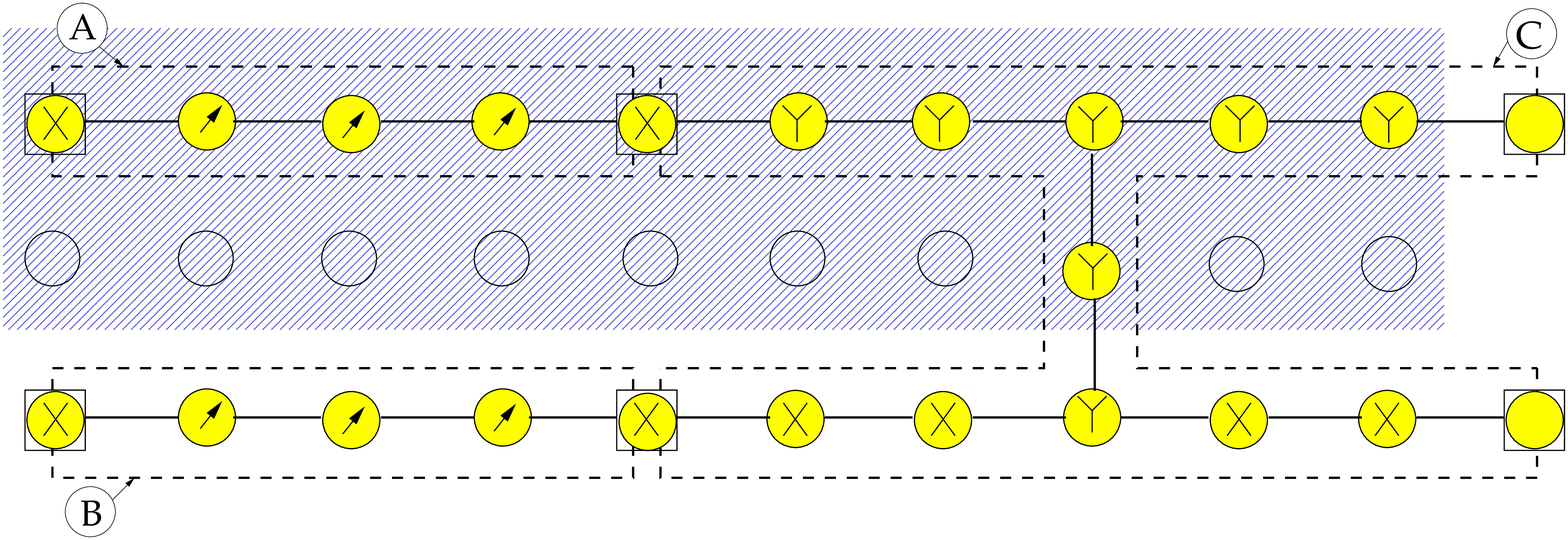,width=7.9cm} 
                            \vspace{0.3cm}\\
                            \footnotesize{(b)} \hspace{0.5cm} \parbox{4.2cm}{
                                                                             \epsfig{file=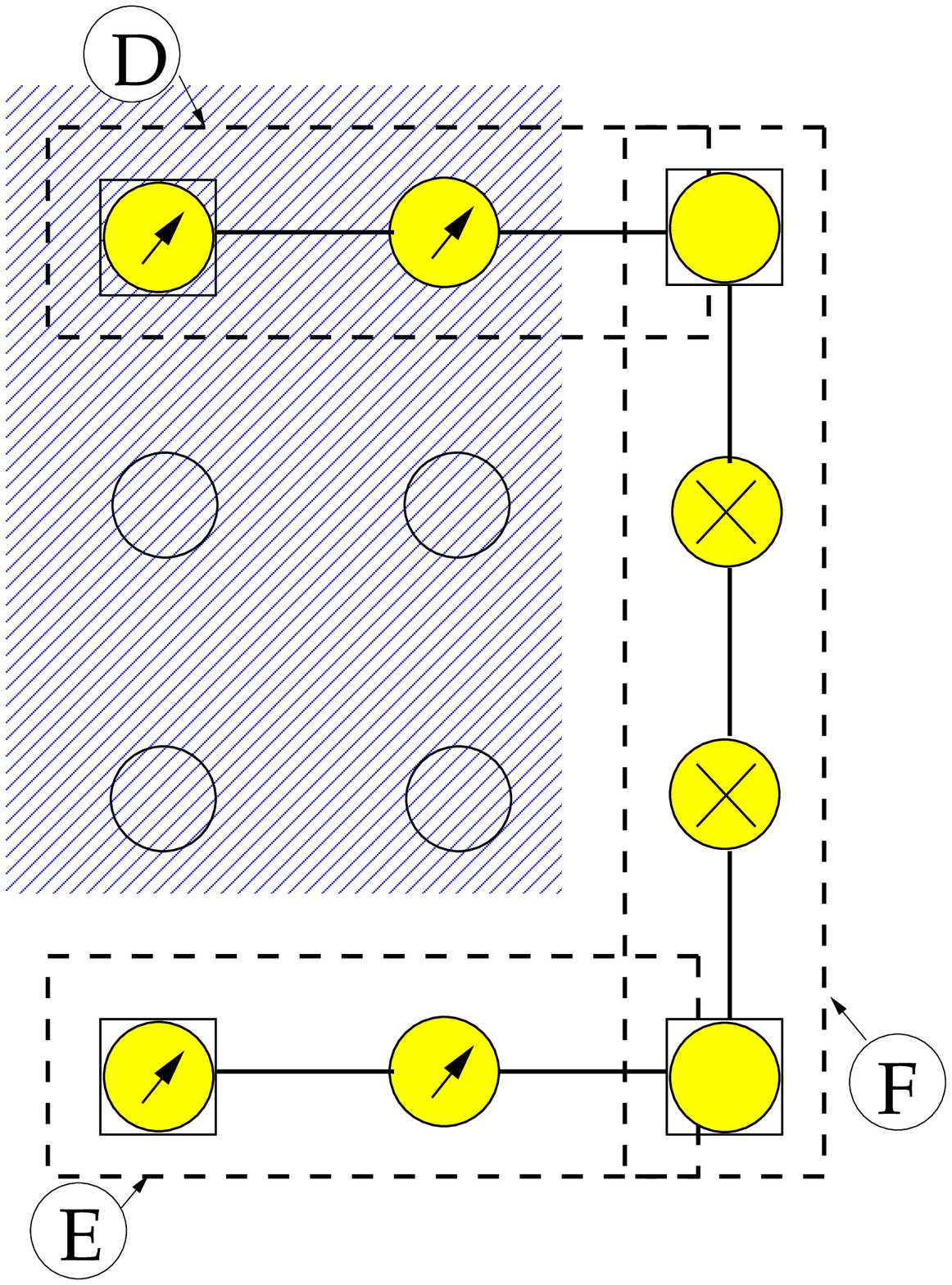,width=2.7cm}
                                                                            }
         \end{tabular}
\caption{\label{1wqc_compare}
         \footnotesize{
                       Dashed boxes A,B realize general one-qubit unitaries and D,E one-qubit unitaries with decomposition $U_x(\theta)U_z(\phi)$. C is a fifteen-qubit realization for the CNOT found in \cite{raussen03} and F is the remote controlled-phase. Arrows indicate measurements in a general direction on the equator of the Bloch sphere. The shaded regions denote the repeated computation units per logical qubit. Here again, empty circles indicate qubits measured in the $Z$-basis.
                      }
         }
     \end{center}
\end{figure}

The first application of the remote-$\Lambda(Z)$ circuits of  Fig.$\,$\ref{remote_C-phase} is a more economic way of doing computation in the 1WQC model using a two-dimensional cluster state. In that case, either a {\sc cnot} pattern using extra teleportation steps for the control state (Fig.$\,$\ref{1wqc_cnot_square}) can be used or alternatively, a pattern with fifteen qubits (Fig.$\,$\ref{1wqc_compare}a) has been proposed. Again, we can use the $\Lambda(Z)$ instead of the {\sc cnot}, replacing the known qubit-costly {\sc cnot} realizations with the remote-$\Lambda(Z)$ pattern derived in Fig.$\,$\ref{remote_C-phase}c, without sacrificing our ability to do universal quantum computation.

The economy in qubits is apparent if we consider a typical computation block consisting of one-qubit unitaries applied to two logical qubits, which then interact through the {\sc cnot} or  alternatively the $\Lambda(Z)$. These one-qubit unitaries can in general be decomposed in terms of the Euler angles $(\phi,\theta,\psi)$ as $U_z(\psi)U_x(\theta)U_z(\phi)$. But, since the rotations around the $z$-axis commute with the controlled-phase gates, it is sufficient to realize one-qubit unitaries of the form $U_x(\theta)U_z(\phi)$ followed by a remote-$\Lambda(Z)$ gate at each repeated unit of the computation. Composing the patterns of Fig.$\,$\ref{cluster gates}a,b and eliminating the intermediate consecutive measurements in the $X$-basis (which realize a redundant quantum wire according to Fig.$\,$\ref{1wqc_tel}), two measured qubits are needed to realize a $U_x(\theta)U_z(\phi)$ rotation, in contrast to the four measured qubits needed to realize a general one-qubit rotation used originally in the 1WQC constructions.

Comparing first the contruction of Fig.$\,$\ref{1wqc_compare}a with the one of Fig.$\,$\ref{1wqc_compare}b, the latter uses less than a third of the total number of qubits and a fifth of the computation length, with an increase of the computation width by one. Computation is performed in a linear fashion from left to right, with the logical qubits separated by regions of qubits measured in the $Z$-basis. The comparison is done by considering the repeated computation units per logical qubit in each case, denoted by the shaded regions in Fig.$\,$\ref{1wqc_compare}. The comparison with the {\sc cnot} pattern of Fig.$\,$\ref{1wqc_cnot_square} can also immediately be made, if we observe that the remote-$\Lambda(Z)$ pattern already forms part of the former in the connection between the third and the ninth qubit. Thus, using the remote-$\Lambda(Z)$ in place of the {\sc cnot} proves more qubit-efficient for both these different {\sc cnot} realizations.

\vspace{0.2cm}

\vspace{0.0cm}
\subsection{Simplifying the TQC model}

Secondly, the remote-$\Lambda(Z)$ circuit of Fig.$\,$\ref{remote_C-phase} suggests a method of affecting a controlled-phase gate between two qubits using measurements of operators of weight at most two. This provides an alternative and greatly simplified proof of the universality of two-qubit measurements in the TQC model, without the need for the preparation of the ancillary state $|a_{\text {\sc cnot}}\rangle$.

Two different measurement procedures are implicit in Fig.$\,$\ref{remote_C-phase}b. The first one, although not minimal in the number of two-qubit measurements used to affect the $\Lambda(Z)$ gate, will be useful as a method for constructing graph states in the following section. Starting with the circuit of Fig.$\,$\ref{remote_C-phase}b, we can interpret each of the boxes A and B as performing an incomplete two-qubit measurement on one half of the entangled state
\[
  |\Omega \rangle = \Lambda(z)|+\rangle |+\rangle = {|+\rangle |0\rangle + |-\rangle |1\rangle \over \sqrt{2}}
\]
and the control and target qubits respectively. The state $|\Omega \rangle$ can also be prepared by a complete two-qubit measurement, giving a total of three two-qubit measurements to affect the controlled-phase gate according to this method.

To elaborate on the exact measurements that will be needed, box A of Fig.$\,$\ref{remote_C-phase}b can be viewed as realizing a measurement of $Z^{(1)}X^{(2)}$ followed by a measurement of the second qubit in the $Z$-basis and similarly for box B. Starting with the state $|\Omega\rangle$, the subsequent measurement procedure should therefore be: (1) measure $ZXII$ and $IIXZ$, then (2) measure $IZII$ and $IIZI$. The analysis for the evolution of the stabilizer generators under the proposed measurement sequence is given in table \ref{table}. To prove that the controlled-phase gate between the first and the fourth qubit is in fact realized, we also follow the evolution of the initial (or logical) $X$,$Z$ operators on those qubits, denoted by $\bar{X},\bar{Z}$, keeping in mind that $\Lambda(Z)$ after conjugation takes $X\otimes I$ to $X\otimes Z$ and $Z\otimes I$ to itself.

\begin{table}[tb]
\[
\small{
  \begin{array}{rrccccccrrccccc}
            & \multicolumn{14}{l}{\text{{\sc At start}}} \vspace{0.1cm}\\
            & S:         & I & X & Z & I & & & & & & & & & \\
            &            & I & Z & X & I & & & & & & & & & \\
            & \bar{X_1}: & X & I & I & I & & & & & & & & & \\
            & \bar{Z_1}: & Z & I & I & I & & & & & & & & & \\
            & \bar{X_4}: & I & I & I & X & & & & & & & & & \\
            & \bar{Z_4}: & I & I & I & Z & & & & & & & & & \\ &  \vspace{0.1cm}\\
            & \multicolumn{6}{c}{\text{1a) {\sc Measure} $ZXII$}} & \hspace{0.2cm}& & \multicolumn{6}{c}{\text{1b) {\sc Measure} $IIXZ$}} \vspace{0.1cm}\\
            & S:         & I & X & Z & I & & & & S:         & I & I & X & Z &\\
            &            & Z & X & I & I & & & &            & Z & X & I & I &\\ 
            & \bar{X_1}: & X & Z & X & I & & & & \bar{X_1}: & X & Z & X & I &\\ 
            & \bar{Z_1}: & Z & I & I & I & & & & \bar{Z_1}: & Z & I & I & I &\\
            & \bar{X_4}: & I & I & I & X & & & & \bar{X_4}: & I & X & Z & X &\\ 
            & \bar{Z_4}: & I & I & I & Z & & & & \bar{Z_4}: & I & I & I & Z &\\ &  \vspace{0.1cm}\\
            & \multicolumn{6}{l}{\text{2a) {\sc Measure} $IZII$}} & & & \multicolumn{6}{l}{\text{2b) {\sc Measure} $IIZI$}} \vspace{0.1cm}\\
            & S:         & I & I & X & Z & & & & S:         & I & I & Z & I &\\
            &            & I & Z & I & I & & & &            & I & Z & I & I &\\
            & \bar{X_1}: & X & Z & X & I & & & & \bar{X_1}: & X & I & I & Z &\\ 
            & \bar{Z_1}: & Z & I & I & I & & & & \bar{Z_1}: & Z & I & I & I &\\
            & \bar{X_4}: & Z & I & Z & X & & & & \bar{X_4}: & Z & I & I & X &\\ 
            & \bar{Z_4}: & I & I & I & Z & & & & \bar{Z_4}: & I & I & I & Z
  \end{array}
     }
\]
\caption{\label{table}
         Verifying that the proposed measurement procedure realizes $\Lambda(Z)_{(1,4)}$. S is the set of stabilizer generators at each step. $\bar{X_1},\bar{Z_1}$ and $\bar{X_4}, \bar{Z_4}$ indicate the logical $X$,$Z$ operators on the first and fourth qubit respectively. 
        }
\end{table}

The measurement procedure with the minimal number of two-qubit measurements is obtained if we interpret the measurement of the second qubit as a measurement of $Z^{(1)}Z^{(3)}$ (box C) and then the measurement on the third qubit as a measurement of $X^{(3)}Z^{(4)}$ (box B) as before. Eliminating the second qubit altogether and renumbering the rest, the measurement procedure is: (1) measure $ZZI$, (2) measure $IXZ$ and (3) measure $IZI$. The last operator to be measured is chosen so that it anticommutes with the previously measured operator and therefore projects out the ancillary second qubit. Although the two measurements $ZZI$ and $IXZ$ do not commute, we note that if the measurements are performed in the opposite order, then the gate is still applied if the ancillary qubit is initialized in the $|0\rangle$ instead of the $|+\rangle$ state and finally projected on the $x$-axis instead of the $z$-axis.

In case the {\sc cnot} is to be realized instead of the controlled-phase, the measurement procedure above can be straightforwardly modified by conjugating the third qubit with a Hadamard, thereby changing the measured operator $IXZ$ to $IXX$. In fact, a method for realizing the {\sc cnot} using just two two-qubit measurements, essentially identical to the one we just presented, has already appeared in the literature in the context of fault-tolerance in higher-dimensional systems \cite{gottes98}, but its relevance to the TQC model had not been hitherto appreciated. A surprising feature of our result in this section is that we have explicitly drawn a connection between the remote-{\sc cnot} circuit of Fig.$\,$\ref{remote_cnot} and the seemingly unrelated method for affecting the {\sc cnot} according to the measurement sequence $IXX$, $ZZI$ and $IXI$, both of which have been proposed for the realization of the {\sc cnot} gate in very different contexts and without reference to one another.

It is interesting to note that only two two-qubit measurements are required to perform the $\Lambda(Z)$ gate according to this second measurement procedure, in contrast to the five two-qubit measurements needed to perform {\sc cnot} in the method illustrated in Fig.$\,$\ref{CNOT}. Moreover, this is the minimal number of two-qubit measurements possible, since in order to simulate a two-qubit gate with measurements each one of the two qubits need to interact at least once with the ancillary state.

\vspace{0.0cm}
\subsection{Constructing graph states with two-qubit measurements}

As discussed in the introduction, the cluster state forms a substrate for universal quantum computation within the 1WQC model. The essense of its universality lies in the observation that, by appropriately deleting qubits, an arbitrary sequence of gates taken from a universal gate set can be imprinted and then executed on it. But, as has been repeatedly hinted on in the literature \cite{raussen03}, this is not necessarily the most beneficial way of thinking about doing computation in the 1WQC model. A powerful indication towards this comes from the surprising fact that all the computational task that lies within the Clifford group can be executed simultaneously in the 1WQC, resulting in a transformation of the initial cluster state to a general graph state (up to local Pauli operators).

Due to this fact and in cases when an application-specific quantum computer would be desirable (we think of factoring, for example) the computation can equally well be performed by circumventing the Clifford part measurements and starting directly with a specific graph state. In addition, this may practically even prove preferable for reasons pertaining to the robustness of the computation: removing altogether the Clifford part measurements will significantly reduce the total number of qubits involved in the computation, thus decreasing the possible locations of error. Such application-specific graph states will then have to be constructed and verified (using graph purification protocols \cite{dur03}) before the computation initiates \cite{priv_comm}.

Since operationally any graph state can be realized by selectively applying the controlled-phase gate between pairs of adjacent qubits initialized in the $|+\rangle$ state, it is interesting to investigate how a given graph state can be constructed within a measurement-based model for performing quantum computation. In particular, in this section we will discuss how the two measurement procedures for affecting the controlled-phase gate, that were inspired by the circuit of Fig.$\,$\ref{remote_C-phase}b, give rise to explicit methods for constructing an arbitrary graph state.

\begin{figure}[tb]
      \begin{center}
         \begin{tabular}{c}
                            \footnotesize{(a)} \vspace{0.2cm} \\
                            \epsfig{file=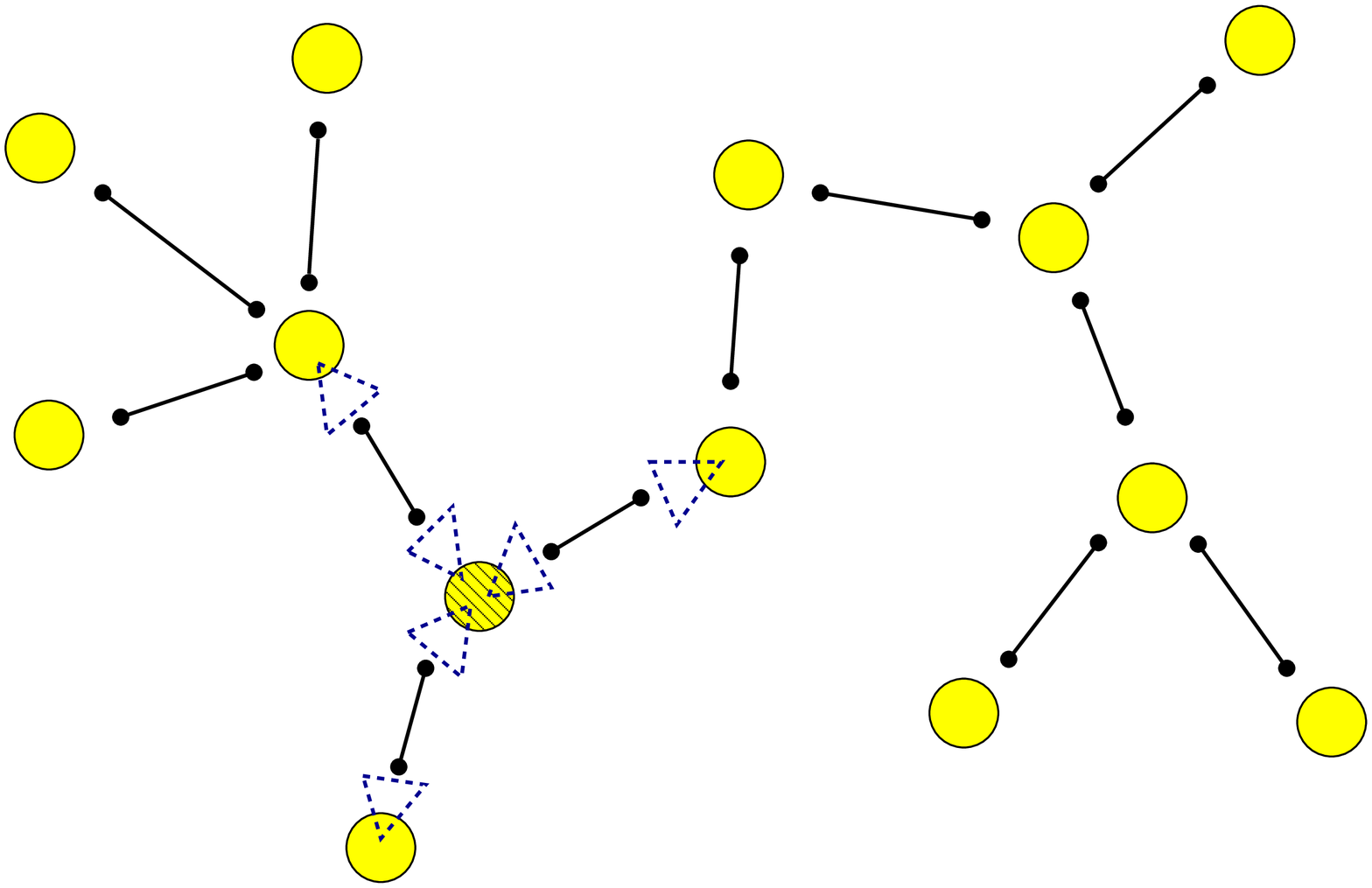,width=7.5cm} \vspace{0.3cm}\\
                            \footnotesize{(b)} \vspace{0.2cm} \\
                            \epsfig{file=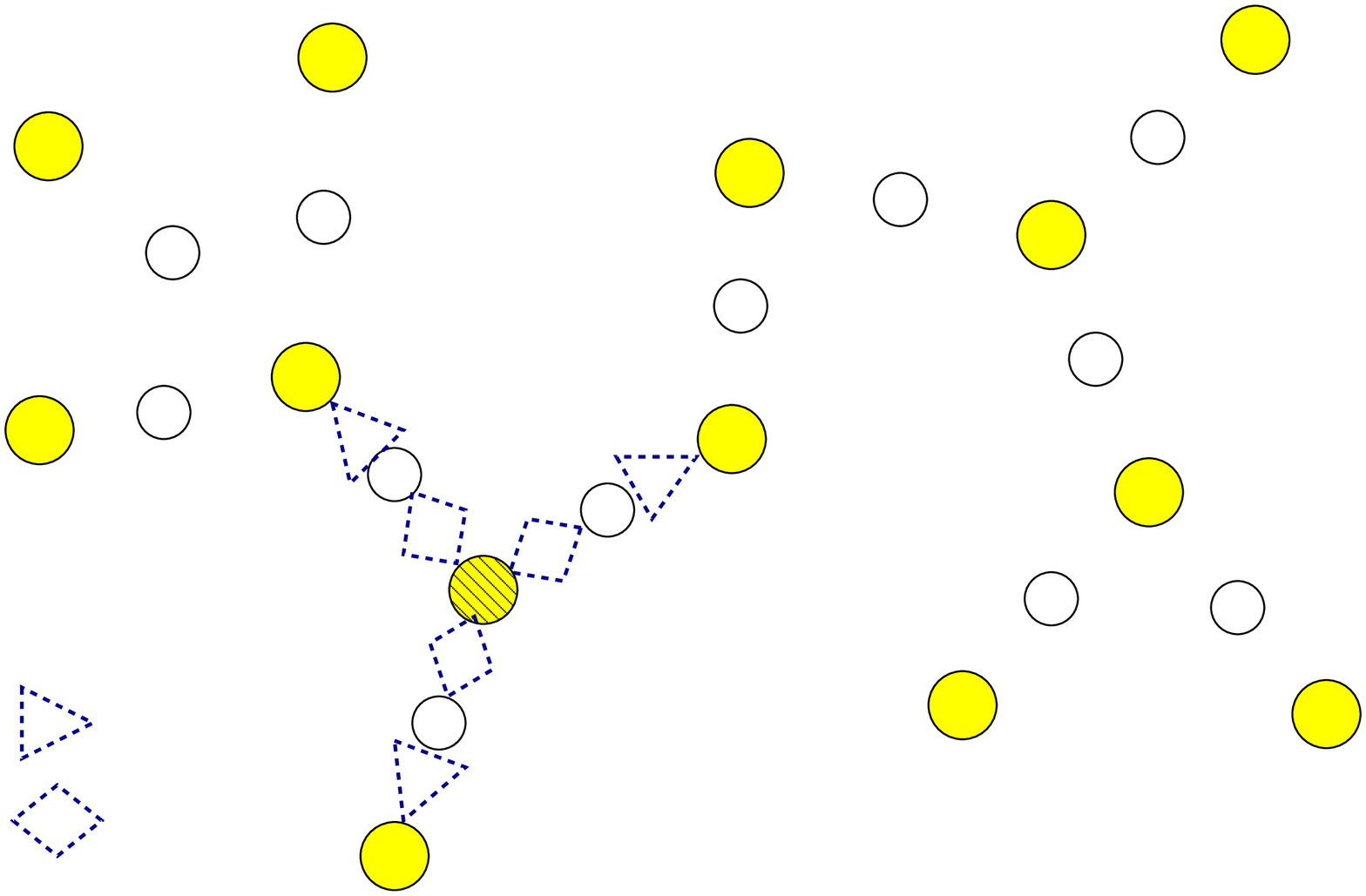,width=7.5cm} \put(-6.92,0.85){:$\,XZ$} \put(-6.92, 0.32){:$\,ZZ$}
              
         \end{tabular}

\caption{\label{graph_prepare}
         \footnotesize{
Creating an arbitrary graph with two-qubit measurements. For clarity only the measurements needed to create the edges incident on the central striped qubit are shown. Filled circles represent qubits of the graph initialized in the $|+ \rangle$ state, and empty circles represent the extra ancillary qubits. The dark dots connected with a bar indicate two qubits in the entangled state $|\Omega \rangle$. Dashed triangles represent measurements of $XZ$ and rhombuses of $ZZ$. All ancillary qubits are simultaneously measured in the final step in appropriate bases.
                      }
         }  
      \end{center}
\end{figure}

The first measurement procedure of the previous section can be used to affect the controlled-phase gate between the qubits in our graph, using as ancillae a number of copies of the state $|\Omega \rangle$ equal to the number of edges created. Implementing the two-qubit measurements needed will require a number of measurement steps at least equal to the vertex degree of the graph under construction, since all measurements with support on a common qubit will have to be realized at different times. Considering that all one-qubit measurements on the ancillary qubits can be performed simultaneously in the final step, the actual logical depth of the graph preparation will equal one plus the maximum vertex degree of the graph to be constructed. An example of the graph state preparation is shown schematically in Fig.$\,$\ref{graph_prepare}a. In order to create all edges connected to the striped qubit of this graph, three two-qubit incomplete measurements, each represented by a dashed triangle, must be done consecutively followed by three one-qubit projections performed simultaneously in the fourth and last step.

At this point, it is intriguing to consider the similarities of this method for constructing graph states to the valence bond solid model for doing quantum computation \cite{verstraete03}. In fact, if the distinction between the qubits of the graph (denoted by filled circles in Fig.$\,$\ref{graph_prepare}) and the qubits in the entangled state $|\Omega \rangle$ (dark dots) is lifted and they are all viewed as physical qubits residing on different sites of a solid, then the correlations between the qubits in the entangled states can be interpreted as arising from valence bonds formed between qubits on neighbouring sites. In such a valence bond model, our viewpoint for constructing the graph state is retained by identifying in each site one of these physical qubits as the logical one and performing appropriate measurements locally on the sites in order to form the desired graph state. The subsequent computation can then be performed by one-qubit measurements, in accordance to the 1WQC model.   

Naturally, the graph state can be constructed even in the absence of the ancillary entangled pairs, according to our second measurement procedure proposed in the previous section. In this case, a number of ancillary qubits equal to the number of edges of the graph will be needed. For each edge, the measurement procedure then consists of two two-qubit measurements each performed between an ancillary qubit and one of the two vertices incident on this edge, followed by a projective measurement on the ancillary qubit.

As with our previous method for constructing the graph state, we would like to inquire into the extent to which this measurement sequence can be parallelized, calculating its logical depth. Due to the fact that the number of two-qubit measurements with support on a given graph qubit equals its degree in the graph, the logical depth of this measurement procedure is at least equal to one plus the maximum vertex degree of the graph under construction. Recall that although the two measurements $ZZI$ and $IXZ$ do not commute, the controlled-phase gate is still applied even if they are performed in the opposite order, provided the ancillary qubit is initialized in the $|0\rangle$ instead of the $|+\rangle$ state and finally measured in the  $X$-basis instead of the $Z$-basis. With this observation in mind, the two-qubit measurement sequence for the construction of the graph $G$ can be thought of as an edge-coloring of the larger graph $G'$ constructed by adding the ancillary qubits to the vertex set and dividing each edge of the graph $G$ into two, each between the originally adjacent vertices and the corresponding ancilla. In the new graph $G'$, edges represent the two-qubit measurements to be performed and the requirement that all two-qubit measurements with support on a given qubit need to be executed in different time steps translates into the requirement that edges incident on a given vertex are colored differently. As a consequence, the logical depth of the measurement procedure will be given by the minimum number of colors to realize the edge-coloring of the graph $G'$. Letting $V(G), E(G)$ denote respectively the vertex and edge set of the graph $G$, the graph $G'$ will then be given by 
\begin{eqnarray*}
  V(G') & = & V(G)\cup \{a_1, a_2,\dots ,a_{|E(G)|}\}\\
  E(G') & = & \{(i,a_k), (a_k, j):(i,j)\in E(G)\},
\end{eqnarray*}
where $k\in \{1,2,\dots, |E(G)|\}$ enumerates the edges of $G$ and $a_k$ is used as a label for the $k^{th}$ ancillary qubit. The minimum number of colors necessary to color each edge of $G'$ so that no two edges incident on the same vertex have the same color is defined as its {\it edge-chromatic number $\chi'(G')$}. As we already remarked, $\chi'(G') \geq \Delta(G')$, where $\Delta(G')$ is the maximum vertex degree of $G'$. For graphs with no multiple edges between the same pair of vertices, a remarkable theorem \cite{vizing64,gupta66} states that $ \chi'(G') \leq \Delta(G') + 1$ and for {\it bipartite} graphs (also called {\it two-colorable}) in particular, the equality $\chi'(G') = \Delta(G')$ has been proven \cite{konig16}. But in our case $G'$ is always bipartite by virtue of the fact that the qubits of the graph $G$ are adjacent to ancillary qubits only. Hence, the logical depth for the construction of the graph $G$ equals its maximum vertex degree \cite{note} plus one (to account for the one-qubit projections of the ancillary qubits done simultaneously in the last step). An example for the use of this measurement procedure is given in  Fig.$\,$\ref{graph_prepare}b.

In both measurement procedures mentioned above, the desired graph state is realized up to Pauli operators that depend on all measurement outcomes. However, preparing the graph state up to local Pauli corrections will not jeopardize the computation later performed on it, since these corrections can be absorbed in the initialization of the classical registers that will be used to buffer the subsequent measurement outcomes \cite{raussen03}.

\vspace{0.0cm}
\section{Conclusion}
The mapping that we derived between the two models for doing computation by measurement is eloquent of the deep structural similarities between them. In fact, the 1WQC model can be thought of as naturally emerging from the TQC once the restriction to one-qubit measurements is imposed. In such a case, the mapping reveals that the interaction between qubits will have to be mediated by a two-qubit unitary, which in the case of the 1WQC is taken to be the controlled-phase. In this respect, our mapping sheds light on the inner workings of the 1WQC, helping us understand its operation on a more intuitive level. Moreover, our mapping explicitly demonstrates the correspondence between the local Pauli corrections in the TQC and 1WQC models, which had been strongly suggestive of an underlying connection between them since the time both models were developed. The mapping thus significantly simplifies the proofs for the operation of the 1WQC circuits that so far always allude to theorems based on the stabilizer formalism \cite{raussen03}. If you would concede that the operation of either one of the 1WQC or TQC is more transparent, then this mapping facilitates the understanding of the other model.  In particular, the less familiar model of the 1WQC is thus rendered less enigmatic and more apt for adaptation towards the ultimate goal of achieving fault-tolerance within this model with a realistically low threshold.

In the last section two methods were proposed for the direct construction of an arbitrary graph state making use of two-qubit measurements, one of which is strongly reminiscent of the valence bond model for quantum computation. Apart from the inherent merit of the simplicity of the remote-$\Lambda(Z)$ gate construction itself, this also led us to propose a new scheme for realizing a universal set of operators within the 1WQC which is more qubit-effective than the known ones, as well as to considerably simplify the proof of the universality of two-qubit measurements in the TQC model. The basic construction emerging within only a few logical steps from the well-known remote-{\sc cnot} circuit is another evidence that the mapping occupying the main body of this paper can be viewed as an attractive mental tool when thinking within the two models that perform computation by measurements for swiftly translating useful structures back and forth between them.
 
\vspace{0.0cm}
\begin{acknowledgments}
We note that alternative but similar explanations of the workings of the 1WQC have been given by Michael Nielsen in unpublished notes \cite{nielsen03}. Peter Shor has also independently developed explanations for the 1WQC operation along similar lines. A systematic construction of one-way quantum computer models starting with the one-bit teleportation scheme \cite{zhou00}, that complements the current work, is investigated by Andrew Childs, Michael Nielsen, and one of us \cite{childs04}. During the final preparation of this work, a paper by Simon Perdrix that uses the one-bit teleportation scheme to further simplify the requirements of the TQC model has also come to our attention \cite{perdrix04}.

We are grateful to Robert Raussendorf for numerous insightful discussions on the workings of the 1WQC and to Ben Toner for many helpful comments and corrections. P.A. and D.L are supported by the US NSF under grant no.$\,$EIA-0086038, and D.L. is also supported by the Richard Tolman Foundation.
\end{acknowledgments}




\begin{thebibliography}{99}
\bibitem{note}
Note that $\Delta(G') = \Delta(G)$ and so  $\chi'(G') = \Delta(G)$ for the bipartite $G'$.
\end{thebibliography}
\end{document}